\documentclass{aa}  % longauth
\usepackage{graphicx}
\usepackage{caption}
\usepackage{txfonts}
\usepackage{booktabs}
\usepackage{color}
\usepackage[dvipsnames]{xcolor}
\usepackage{natbib,twoopt}
\usepackage{lscape}
\usepackage{caption}
\usepackage{subcaption}
\usepackage[symbol]{footmisc} % to add footnotes with symbols
\usepackage{tablefootnote}
\usepackage{soul}
\usepackage[bookmarks=true, pdfnewwindow=true, colorlinks=true, allcolors=blue, breaklinks=true, final=true]{hyperref}
\usepackage{orcidlink}
\bibpunct{(}{)}{;}{a}{}{,} % to follow the A&A style
\newcommand{\Nthresh}{\mbox{N$_{\rm threshold}$}}
\newcommand{\av}{\mbox{$A_{\rm V}$}}

\def\purple#1 {{\textcolor{purple}{#1}}\ }
\def\red#1 {\textcolor{red}{#1}}
\def\blue#1 {{\textcolor{blue}{#1}}\ }
\def\zy#1 {\textcolor{forestgreen}{ zy: #1}}
\def\Htwo { H$_2$ }
\def\cmt { cm$^{-3}$ }
\def\Msgb { $M_{\rm sgb}$ }
\def\reference#1 {\textcolor{red}{ reference: #1}}

\begin{document}

\title{Gravitationally Bound Gas Determines Star Formation in the Galaxy}

 %  \subtitle{I. Overviewing the $\kappa$-mechanism}

   \author{Sihan Jiao
          \inst{1\ast}
          \and
          Jingwen Wu\inst{2,1\ast}
%          \fnmsep\thanks{Just to show the usage of the elements in the author field}
          \and
            Zhi-Yu Zhang
          \inst{3,4}
          \and 
          Neal J. Evans II
           \inst{5}
           \and
          Chao-Wei Tsai
           \inst{1,6,2}
                     \and
          Di Li
           \inst{7,1,8}
                     \and
         Hauyu Baobab Liu
           \inst{9,10}
                     \and
        Yong Shi
           \inst{3,4}
                     \and
         Junzhi Wang
           \inst{11}
                     \and
         Qizhou Zhang
           \inst{12}
                     \and
         Yuxin Lin
           \inst{13}
                     \and
         Linjing Feng
           \inst{1,2}
                     \and
         Xing Lu
           \inst{14}
                     \and
         Yan Sun
             \inst{15}
                     \and
         Hao Ruan
           \inst{2,1}
                     \and
         Fangyuan Deng
           \inst{2,1}
}

   \institute{National Astronomical Observatories, Chinese Academy of Sciences, Beijing 100101, China\\
              \email{sihanjiao@nao.cas.cn}
         \and
             University of Chinese Academy of Sciences, Beijing 100049, China. \\
             \email{jingwen@ucas.ac.cn}
%             \thanks{The university of heaven temporarily does not
%                    accept e-mails}
         \and
             School of Astronomy and Space Science, Nanjing University, Nanjing 210093, China
         \and
             Key Laboratory of Modern Astronomy and Astrophysics, Ministry of Education, Nanjing 210093, China
         \and
             Department of Astronomy, The University of Texas at Austin, 2515 Speedway, Stop C1400, Austin, 78712-1205, USA
         \and
             Institute for Frontiers in Astronomy and Astrophysics, Beijing Normal University,  Beijing 102206, China
         \and
             New Cornerstone Science Laboratory, Department of Astronomy, Tsinghua University, Beijing 100084, China
             \and
             Zhejiang Lab, Hangzhou, 311121, China
             \and
             Physics Department, National Sun Yat-Sen University, Kaohsiung City 80424, Taiwan
             \and
             Center of Astronomy and Gravitation, National Taiwan Normal University, Taipei 116, Taiwan
             \and
             School of Physical Science and Technology, Guangxi University, Nanning 530004, China
             \and
             Harvard-Smithsonian Center for Astrophysics, 60 Garden Street, Cambridge, 02138, USA
             \and
             Centre for Astrochemical Studies, Max-Planck-Institut f\"{u}r Extraterrestrische Physik, Gie{\ss}enbachstra{\ss}e 1, 85748 Garching, Germany
             \and
             Shanghai Astronomical Observatory, Chinese Academy of Sciences, Shanghai 200030, China
             \and
             Purple Mountain Observatory, Chinese Academy of Sciences, Nanjing 210023, China
             \and 
             $^{\ast}$ These authors contributed equally
}

%   \date{Received September 15, 1996; accepted March 16, 1997}

% \abstract{}{}{}{}{} 
% 5 {} token are mandatory
 
  \abstract
 { 
   Stars form from molecular gas under complex conditions influenced by multiple competing physical mechanisms, such as gravity, turbulence, and magnetic fields. However, accurately identifying the fraction of gas actively involved in star formation remains challenging. Using dust continuum observations from the {\it Herschel} Space Observatory, we derived column density maps and their associated probability distribution functions (N-PDFs). Assuming the power-law component in the N-PDFs corresponds to gravitationally bound (and thus star-forming) gas, we analyzed a diverse sample of molecular clouds spanning a wide range of mass and turbulence conditions. This sample included 21 molecular clouds from the solar neighborhood ($d<$500 pc) and 16 high-mass star-forming molecular clouds. For these two groups, we employed the counts of young stellar objects (YSOs) and mid-/far-infrared luminosities as proxies for star formation rates (SFR), respectively.  Both groups revealed a tight linear correlation between the mass of gravitationally bound gas and the SFR, suggesting a universally constant star formation efficiency in the gravitationally bound gas phase. The star-forming gas mass derived from threshold column densities ($N_{{\mbox{\scriptsize threshold}}}$) varies from cloud to cloud and is widely distributed over the range of $\sim$1--17$\times$10$^{21}$ cm$^{-2}$ based on N-PDF analysis. But in solar neighborhood clouds, it is in rough consistency with the traditional approach using \av $\ge$ 8 mag. In contrast, in high turbulent regions (e.g., the Central Molecular Zone) where the classical approach fails, the gravitationally bound gas mass and SFR still follow the same correlation as other high-mass star-forming regions in the Milky Way. Our findings also strongly support the interpretation that gas in the power-law component of the N-PDF is undergoing self-gravitational collapse to form stars. }


\maketitle
%
%-------------------------------------------------------------------

\section{Introduction}

Stars form out of interstellar gas when gravity overwhelms pressure, turbulence, and magnetic fields \citep{Elmegreen1989}.  
In both the Milky Way and external galaxies, most stars form in dense regions in molecular clouds \citep[e.g.,][]{Gao2004b,Gao2004a,Wu2005,Wu2010,Heyer2016AA,Hu2022MNRAS,Neumann2023MNRAS}.

However, how stars form under the competition between gravity and other factors, and what controls the rate and efficiency of star formation in molecular clouds, remain unclear.  

Connections between gas content and star formation rate (SFR) have been established through empirical studies, such as the Kennicutt-Schmidt relation \citep{Schmidt1959,Kennicutt1998,KE2012}.
%for the total neutral gas,
%although with obvious diversities in the efficiency of star formation in normal and starburst galaxies\cite{Gao2004a,Gao2004b}.
These works show diverse efficiencies of star formation between normal and starburst galaxies \citep{Gao2004b,Gao2004a}, indicating different star formation processes. 
The dense gas phase, i.e., with a threshold of \Htwo\ number density above $10^4$\cmt\ or a threshold of optical extinction \av $\ge$ 8 mag, was found to better correlate with SFR  over a wide range of scales, both within the Milky Way \citep[e.g.,][]{Wu2005,Wu2010,Heyer2016AA,Stephens2016ApJ,Pokhrel2021ApJ,Pokhrel2021ApJ,Hu2022MNRAS} and in extragalactic systems \citep[e.g.,][]{Gao2004a,Gao2004b,Zhang2014,Gallagher2018ApJ,Jimenez2019ApJ,Bemis2019AJ,Neumann2023MNRAS}.

For example, a roughly linear correlation was found between the SFR and the dense gas mass by taking an empirical column density threshold of \av = 8 mag, for nearby star forming clouds \citep[e.g.,][]{lada2010,Evans2014}.   

However, distinct exceptions still exist: very turbulent clouds near galactic centers,  such as the central molecular zone (CMZ) of the Milky Way, contain a large amount of dense gas, but the observed SFRs are lower than predicted by an order of magnitude \citep[e.g.,][]{Longmore2013}.  
Therefore, a simple threshold of \av = 8 mag does not adequately distinguish the star-forming gas in molecular clouds. 

Recently, it has been recognized that much of the molecular gas is not gravitationally bound (hereafter ``unbound'') \citep[e.g.,][]{MD2017, Evans2021}.
Simulations of such gas indicate very low star formation rates \citep{Kim2021}.
%The bound gas, although containing a smaller amount of mass, contains the material directly relevant to star formation. 
The bound gas, despite comprising a smaller fraction of the total mass, contains the material directly relevant to star formation.
%Accounting for the lower
Considering the low efficiency in unbound clouds and the metallicity-dependence of cloud mass estimates \citep{Gong2020ApJ...903..142G,Hu2022ApJ...931...28H}, the bound gas produces a much lower SFR than that predicted by simple free-fall models, providing a solution to the low star formation efficiency problem of the Milky Way \citep{Evans2022} \footnote{The SFR from free-fall prediction is 150 to 180 times larger than observed values \citep{Zuckerman1974,McKee2007,Evans2021}.}.

However, precise measurement of the mass of bound gas within a molecular cloud remains challenging, primarily due to the intricate interplay between gravity and turbulence across various scales. 
To address this challenge, the concept of ``dense gas mass" has been introduced, typically inferred from the line luminosity of molecular transitions with high critical densities ($n_{\rm crit}> 10^4$\cmt), such as HCN and CS transitions \citep{Wu2010,Zhang2014,Liutie2016ApJ}.  
This method involves a complex and uncertain conversion from line luminosity to the mass of dense gas. 
For example, a large fraction of HCN 1-0 emission is found to be actually generated in extended, diffused gas regions rather than dense clumps in molecular clouds \citep{Stephens2016ApJ,Evans2020}. 
In addition, these tracers may not necessarily always trace the bound gas, especially in regions with unusually high turbulence. 
Given that different dense gas tracers may give different conversions from line luminosity to dense gas mass, all of these make the accurate estimation of the mass of star forming gas very impractical. 
%in unusually turbulent regions. 

The column density probability distribution function (N-PDF) was proposed as a powerful tool to quantify gas components. 
The turbulent, low-density gases would appear as a lognormal distribution \citep{Federrath2010,Kainulainen2014Sci...344..183K,Federrath2016}, powered by the atomic gas around GMCs \citep{Burkhart2015}.  
At higher densities, the N-PDF develops a power-law tail, generated by self-gravitating gas \citep{Klessen2000,Burkhart2017}.  
%The breakpoint, \Nthresh, between the lognormal and power-law profiles divides the cloud mass between unbound and bound portions.
The breakpoint, \Nthresh, distinguishing between the lognormal and power-law profiles, presents the division of cloud mass structure into unbound and bound portions. 

The N-PDF can be observationally measured both from molecular lines like CO \citep[e.g.,][]{Schneide2016A&A,Orkisz2017A&A} or from dust absorption or emission.  
Dust based N-PDF benefits from being free of opacity and depletion problems that seriously constrain the use of CO for N-PDF, especially in dense regions. 
{\it Herschel} \citep{Pilbratt2010A&A} has enabled the N-PDF study with far-IR dust emission at good resolution and sensitivity, towards quite a few individual star forming regions \citep[e.g.,][]{Schneider2013,lombardi2015,chen2018,Schneider2022}.

Taking advantage of its low optical depth, in this work we use dust emission to generate the N-PDF in a representative sample of nearby low-mass and distant, massive molecular clouds, to separate  bound gas and unbound gas, and to quantitatively measure the mass of  bound gas. 
We also test if gravitational bound gas is a good representative of star forming gas. 
Then we make further tests in molecular clouds in the CMZ, to check if it can explain the low star formation rate in the CMZ. 
All the clouds have high sensitivity far-infrared images of dust emission, observed with  {\it Herschel} space telescope, using PACS \citep{Poglitsch2010} and SPIRE \citep{Griffin2010}. 
In Section \ref{sec:sample}, we describe the sample selection and the methods we used to generate N-PDF for these sources. 
The major results including correlations between the derived gravitational bound gas mass and star formation rates are presented in Section \ref{sec:res}. 
In Section \ref{sec:dis} we discuss the factors that determine the rate and efficiency of star formation in the Galaxy. 
A conclusion is given in Section \ref{sec:con}. 

%--------------------------------------------------------------------
\section{Sample selection and Methods}\label{sec:sample}

\subsection{Sample selection}

To study the correlation between self-gravitating gas masses and star formation rates over a wide dynamic range, we selected a sample of nearby star-forming regions as well as distant and massive clouds \citep{lada2010,Wu2010,Evans2014}.
This sample encompasses molecular clouds with masses ranging from 10$^{2}$ to 10$^{5}$ $M_{\odot}$.

\subsubsection{Low-mass star-forming regions} 

The sample of low-mass star-forming regions combines data from references \cite{lada2010} and \cite{Evans2014}. 
The first subsample \citep[e.g.,][]{lada2010} includes 11 clouds from eight nearby star-forming regions within a distance of less than 500 pc. 
The 2MASS images were adopted to estimate the extinction for calculating gas mass above $\rm A_V =8$ mag. 
The second subsample \citep[e.g.,][]{Evans2014} consists of 29 clouds from 12 star-forming regions. 
Both {\it Spitzer} ( 3.6 to 160 $\mu$m bands) and 2MASS data were utilized to generate extinction maps for estimating the gas mass above $\rm A_{V} = 8$ mag. 
Five star-forming regions (Lupus I, II, II, Perseus, and Ophiuchus) are contained in both subsamples.
The differences between the derived gas masses above $\rm A_{V} = 8$ magnitudes from the different subsamples are less than 50\%.
For target regions not covered by \cite{Evans2014}, we adopted gas mass estimates above $\rm A_{V} = 8$ magnitudes from \cite{lada2010}.
Utilizing the YSO counting approach to calculate the star formation rate, the combination of these two subsamples provides the best available nearby star-forming cloud sample with good SFR estimation in the literature to study SFR-related correlations. \footnote{https://doi.org/10.34515/CATALOG.HINODE-00000}.

To ensure uniform {\it Herschel} data quality across our sample, we searched the combined list in the {\it Herschel} Gould Belt survey (HGBS) archive, where we found deep PACS and SPIRE data for 21 clouds. 
%\textcolor{red}{Baobab: There are 21 clouds in Table 1?}
Detailed descriptions of the observations and data reductions can be found in Section \ref{method:data}.
We adopted the distance and star formation rate (SFR) calculated from YSO counting as reported in \cite{lada2010,Evans2014}, and \cite{Zucker2020}.  
The resulting list of 20 clouds is presented in Table \ref{tab:1}.

%%%%%%%%%%%%%%%%% table 1 %%%%%%%%%%%%%%%%%%%
\begin{table*}[!th]
%\linespread{1.3}
%\rmfamily\\
\centering
\caption{
Basic information for nearby star-forming clouds.
}
\begin{tabular}{llllllllll}\hline
        Clouds         & Distance & $M_{\rm A_V>8}^{\rm 2MASS}$ & $M_{\rm A_V>8}^{\rm Herschel}$    & $M_{\mbox{\scriptsize sgb}}$ & SFR                               & $N_{{\mbox{\scriptsize threshold}}}$  \\
               & (pc)     & ($M_{\odot}$) & ($M_{\odot}$)                         & ($M_{\odot}$)                & (10$^{-6}$ $M_{\odot}$ yr$^{-1}$) & (cm$^{-2}$)                           &
\\\hline\hline
Aquila         & 278$\pm{13}$      & 16034$\pm{2405}$ & 4993$\pm{749}$                                 & 6434$^{+1286}_{-2573}$                       & 322.3$\pm{9.0}$                             & $6.69^{+0.10}_{-0.04}  \times   10^{21}$  \\
California     & 454$\pm{22}$      & 3199$\pm{480}$ & 629$\pm{94}$                                 & 6153$^{+1259}_{-1251}$                         & 70.0$\pm{4.2}$                              & $1.84^{+0.07}_{-0.10}  \times   10^{21}$   \\
Cepheus I      & 336$\pm{16}$      & 6$\pm{1}$ & 209$\pm{31}$                                   & 715$^{+99}_{-71}$                          & 8.5$\pm{1.5}$                              & $1.35^{+0.06}_{-0.05}  \times   10^{21}$  \\
Cepheus II     & 337$\pm{16}$      & 12$\pm{2}$ & 41$\pm{6}$                                    & 86$^{+10}_{-14}$           & 0.0                            & $2.79^{+0.23}_{-0.04} \times   10^{21}$  \\
Cepheus III    & 341$\pm{17}$      & 41$\pm{6}$  & 42$\pm{7}$                                    & 319$^{+48}_{-16}$                          & 10.5$\pm{1.6}$                              & $1.6 \times   10^{21}$                 \\
Cepheus IV     & 359$\pm{17}$      & 0   & 50$\pm{8}$                                     & 4$^{+1}_{-1}$                          & 0.5$\pm{0.4}$                             & $17.47^{+5.67}_{-2.83} \times   10^{21}$   \\
Cepheus V      & 364$\pm{18}$      & 31$\pm{5}$  & 82$\pm{12}$                                   & 1047$^{+244}_{-185}$                          & 4.8$\pm{1.1}$                              & $0.97^{+0.04}_{-0.05} \times   10^{21}$   \\
Cham I   & 210$\pm{14}$      & 176$\pm{26}$ & 263$\pm{39}$                                   & 477$^{+52}_{-47}$                          & 20.5$\pm{2.3}$                             & $2.94^{+0.03}_{-0.03} \times   10^{21}$   \\
Cham II        & 190$\pm{13}$      & 64$\pm{10}$   & 17$\pm{3}$                                    & 344$^{+52}_{-24}$                          & 6.0$\pm{1.1}$                             & $3.2 \times   10^{21}$                 \\
Cham III       & 161$\pm{11}$      & 0 & 12$\pm{2}$                                    & 3$^{+1}_{-1}$                          & 1.0$\pm{0.5}$                             & $10.80^{+1.05}_{-1.61} \times   10^{21}$   \\
Lupus I        & 151$\pm{10}$      & 75$\pm{8}$ & 19$\pm{3}$                                    & 245$^{+37}$                          & 3.3$\pm{0.9}$                              & $1.1 \times   10^{21}$                \\
Lupus III      & 197$\pm{13}$      & 163$\pm{24}$   & 24$\pm{4}$                                   & 212$^{+51}_{-51}$                          & 17.0$\pm{2.1}$                              & $1.61^{+0.04}_{-0.04} \times   10^{21}$   \\
Lupus IV       & 151$\pm{10}$      & 124$\pm{19}$ & 19$\pm{3}$                                   & 453$^{+133}_{-104}$                         & 3.0$\pm{0.9}$                              & $0.75^{+0.01}_{-0.01}  \times   10^{21}$  \\
Musca          & 160$\pm{13}$     & 0 & 4$\pm{1}$                                    & 253$^{+44}_{-41}$                          & 3.0$\pm{0.9}$                             & $1.58^{+0.02}_{-0.02}  \times   10^{21}$  \\
Ophiuchus      & 128$\pm{6}$      & 1209$\pm{181}$ & 466$\pm{70}$                                  & 963$^{+352}_{-201}$                          & 72.8$\pm{4.3}$                             & $3.25^{+0.05}_{-0.07} \times   10^{21}$   \\
Orion A        & 399$\pm{19}$      & 13721$\pm{2058}$ & 10205$\pm{1531}$                                 & 26579$^{+4781}_{-9134}$                        & 715.5$\pm{13.4}$                            & $2.05^{+1.95}_{-0.16} \times   10^{21}$   \\
Orion B        & 415$\pm{22}$      & 7261$\pm{1074}$ & 3488$\pm{523}$                                  & 8617$^{+690}_{-602}$                         & 158.8$\pm{6.3}$                            & $2.95^{+0.14}_{-0.14}  \times   10^{21}$  \\
Perseus        & 256$\pm{12}$      & 1880$\pm{240}$ & 986$\pm{148}$                                  & 4754$^{+1178}_{-991}$                         & 149.5$\pm{6.1}$                            & $1.50^{+0.01}_{-0.01} \times   10^{21}$   \\
Pipe           & 180$\pm{9}$      & 178$\pm{27}$ & 30$\pm{5}$                                   & 122$^{+42}_{-24}$                           & 5.3$\pm{1.2}$                              & $4.18^{+0.01}_{-0.01} \times   10^{21}$   \\
Serpens        & 425$\pm{21}$      & 4213$\pm{632}$ & 1056$\pm{158}$                                  & 4482$^{+2007}_{-1514}$                         & 56.0$\pm{3.7}$                            & $3.21^{+0.26}_{-0.30} \times   10^{21}$   \\
Taurus         & 156$\pm{7}$      & 1766$\pm{265}$ & 402$\pm{60}$                                 & 3250$^{+652}_{-652}$                         & 83.8$\pm{4.6}$                             & $1.62^{+0.02}_{-0.02} \times   10^{21}$  \\
\hline
\end{tabular}
\label{tab:1}
\tablefoot{
The masses derived from $A_{\mbox{\scriptsize V}}$ extinction, and star formation rates are from References: \cite{lada2010} and \cite{Evans2014}. 
The distance is from \cite{Zucker2020}. 
$M_{\rm A_V>8}^{\rm 2MASS}$ is the dense gas mass within the extinction contour of $A_{\mbox{\scriptsize V}}$ = 8 mag, which is from \cite{lada2010} and \cite{Evans2014}; $M_{\rm A_V>8}^{\rm Herschel}$ is the dense gas mass above $N(\rm H_{2})$ = 7.4$\times$10$^{21}$ cm$^{-2}$ from {\it Herschel} column density maps.
$N_{{\mbox{\scriptsize threshold}}}$ is the threshold column density of the turning position between lognormal and power-law distribution in N-PDF, and the corresponding errors are from N-PDF fitting.
\Msgb is the gas mass above $N_{{\mbox{\scriptsize threshold}}}$.
}
\end{table*}
%%%%%%%%%%%%%%%%% table 1 %%%%%%%%%%%%%%%%%%%

\subsubsection{High-mass star-forming regions} 

We selected high-mass star-forming clouds from \cite{Wu2010}, which is a well-studied massive star-forming clump sample.
The sources mapped in CS 5-4 \citep{Shirley2003ApJS..149..375S} have virial masses within the nominal core radius ranging from 30 $M_{\odot}$ to 2750 $M_{\odot}$, with a mean of 920 $M_{\odot}$. 
Most sources in this category contain compact or ultracompact H II regions.
%have infrared luminosities ranging from 10$^{3}$ to 10$^{7}$  $L_{\odot}$ and

%%%%%%%%%%%%%%%%% table 2 %%%%%%%%%%%%%%%%%%%
\begin{table*}[!th]
\centering
\caption{
Information and derived parameters for massive star-forming regions.
}
\begin{tabular}{llllll}\hline
Clouds                      & Distance &$M_{\rm A_V>8}^{\rm Herschel}$  & $M_{\mbox{\scriptsize sgb}}$ & SFR                               & $N_{{\mbox{\scriptsize threshold}}}$  \\
                            & (kpc)     & ($M_{\odot}$)          & ($M_{\odot}$)      & (10$^{-6}$ $M_{\odot}$ yr$^{-1}$) & (cm$^{-2}$) \\
               \hline\hline
W3(OH)                      & 2.0$\pm{0.1}$     & 1.8$\pm{0.5} \times   10^{3}$      & $3.9^{+1.1}_{-1.1} \times   10^{3}$        & 18.2 $\pm{9.1}$                            & $3.53^{+0.08}_{-0.01} \times   10^{21}$ \\
G9.62$+$0.10                & 5.2$\pm{0.5}$ & 3.1$\pm{0.9} \times   10^{5}$        & $2.0^{+0.9}_{-0.6} \times   10^{4}$        & 68.0$\pm{34.0}$                             & $1.31^{+0.12}_{-0.23}  \times   10^{22}$ \\
G10.30$-$0.10            & 3.2$\pm{0.1}$   & 4.3$\pm{1.3} \times   10^{5}$   & $1.3^{+0.4}_{-0.3} \times   10^{4}$        & 26.4$\pm{13.2}$                            & $2.62^{+0.57}_{-0.23}  \times   10^{22}$ \\
G10.60$-$0.40               & 5.0$\pm{0.5}$  & 9.4$\pm{2.9} \times   10^{4}$     & $2.3^{+1.1}_{-0.7} \times   10^{4}$        & 145.2$\pm{72.6}$                            & $1.87^{+0.12}_{-0.13}  \times   10^{22}$ \\
G12.89$+$0.49               & 2.5$\pm{0.3}$  & 1.4$\pm{0.4} \times   10^{4}$     & $3.8^{+1.7}_{-1.7} \times   10^{3}$        & 5.2$\pm{2.6}$                              & $1.14^{+0.23}_{-0.22}  \times   10^{22}$ \\
W33A                        & 4.5$\pm{0.4}$   & 1.5$\pm{0.5} \times   10^{3}$    & $5.9^{+2.5}_{-3.0} \times   10^{3}$        & 34.6$\pm{17.3}$                             & $3.60^{+0.09}_{-0.10} \times   10^{22}$ \\
W43M                        & 5.3$\pm{0.5}$  & 1.7$\pm{0.5} \times   10^{5}$     & $9.2^{+2.6}_{-2.1} \times   10^{4}$        & 148.6$\pm{74.3}$                            & $2.61^{+0.02}_{-0.02} \times   10^{22}$ \\
G35.20$-$0.74               & 2.2$\pm{0.2}$  & 1.4$\pm{0.4} \times   10^{4}$      & $1.0^{+0.3}_{-0.3} \times   10^{4}$        & 4.4$\pm{2.2}$                              & $9.35^{+1.89}_{-0.42} \times   10^{21}$ \\
W49N                        & 11.1$\pm{0.9}$  & 5.2$\pm{1.6} \times   10^{5}$     & $2.4^{+0.6}_{-0.9} \times   10^{5}$        & 704.2$\pm{352.1}$                            & $1.10^{+0.08}_{-0.14} \times   10^{21}$ \\
OH43.80$-$0.13              & 6.9$\pm{0.2}$   & 5.1$\pm{1.6} \times   10^{4}$     & $2.0^{+0.4}_{-0.4} \times   10^{4}$        & 35.9$\pm{18.0}$                             & $8.53^{+0.24}_{-0.26}  \times   10^{21}$ \\
W51M                        & 5.4$\pm{0.3}$   & 3.8$\pm{1.1} \times   10^{4}$   & $8.2^{+1.6}_{-1.6} \times   10^{4}$        & 529.5$\pm{264.7}$                            & $2.07^{+0.12}_{-0.24}  \times   10^{22}$ \\
G59.78+0.06                 & 2.2$\pm{0.1}$    & 8.6$\pm{2.6} \times   10^{2}$     & $1.9^{+0.5}_{-0.7} \times   10^{3}$        & 2.7$\pm{1.4}$                              & $5.86^{+0.01}_{-0.09} \times   10^{21}$ \\
S106                        & 4.1$\pm{0.4}$     & 8.3$\pm{2.5} \times   10^{3}$   & $1.1^{+0.6}_{-0.4} \times   10^{4}$        & 114.4$\pm{57.2}$                             & $6.07^{+0.22}_{-0.16} \times   10^{21}$ \\
W75N                        & 1.3$\pm{0.1}$    & 1.9$\pm{0.6} \times   10^{3}$    & $1.7^{+0.4}_{-0.4} \times   10^{3}$        & 8.1$\pm{4.1}$                              & $7.94^{+0.77}_{-1.18}  \times   10^{21}$ \\
DR21 S                      & 1.5$\pm{0.1}$   & 5.6$\pm{1.7} \times   10^{3}$   & $7.2^{+1.4}_{-1.4} \times   10^{3}$        & 30.8$\pm{15.4}$                             & $4.65^{+0.09}_{-0.11} \times   10^{21}$ \\
S158                        & 2.8$\pm{0.2}$   & 1.7$\pm{0.5} \times   10^{4}$    & $2.0^{+0.3}_{-0.4} \times   10^{4}$        & 43.5$\pm{21.8}$                             & $4.77^{+0.11}_{-0.14} \times   10^{21}$ \\
\hline
\end{tabular}
\label{tab:2}
\tablefoot{
The distances are calculated based on parallax measurements from \cite{Reid2014}. 
$N_{{\mbox{\scriptsize threshold}}}$ is the threshold column density corresponding to the turning position between lognormal and power-law distribution in N-PDF. 
 $M_{\rm A_V>8}^{\rm Herschel}$ is the dense gas mass above $N(\rm H_{2})$ = 7.4$\times$10$^{21}$ cm$^{-2}$.
$M_{\mbox{\scriptsize sgb}}$ is the gas mass above $N_{{\mbox{\scriptsize threshold}}}$. 
SFRs are calculated from bolometric luminosity.
%but multiplied by 7.8 to match the SFR from YSO counting.
}
\end{table*}
%%%%%%%%%%%%%%%%% table 2 %%%%%%%%%%%%%%%%%%%

Accurate distance measurements are crucial for calculating the mass of each cloud. 
We conducted a cross-match of this sample with the data from \cite{Reid2014} to identify a subsample for which distances have been accurately determined using parallax measurements. 
Subsequently, we searched the cross-matched sample in the {\it Herschel} archive to select sources with available deep SPIRE and PACS images.
In the resulting subsample, a few sources have low bolometric luminosity.
It has been noticed that for sources with low bolometric luminosity \citep[less than 10$^{4.5} L_{\odot}$,][]{Wu2010} or low SFR \cite[less than 5 $\times 10^{-6}$ $M_{\odot}$ yr$^{-1}$,][]{vutisalchavakul2016}, the bolometric luminosity may no longer well trace real SFR due to the lack of high-mass stars in their IMF.
We therefore remove such targets from our sample.  
The finally selected 16 massive star-forming regions are listed in Table \ref{tab:2}.  

\subsection{Calculating column density}

\subsubsection{Data of dust emission}\label{method:data}

%Dust emission within molecular clouds at the cloud scale is optically thin at submillimeter bands. 
We employed {\it Herschel} \footnote{{\it Herschel} is an ESA space observatory with science instruments provided by European-led Principal Investigator consortia and with important participation from NASA.} data to generate the column density maps of the target molecular clouds. 

For the nearby star-forming regions, we retrieved the HGBS data that were taken at 70/160 $\mu$m using the PACS instrument \citep{Poglitsch2010} and at 250/350/500 $\mu$m using the SPIRE instrument \citep{Griffin2010}.
The HGBS took a census in the nearby (0.5 kpc) molecular cloud complexes for an extensive imaging survey of the densest portions of the Gould Belt, down to a 5 $\sigma$ column sensitivity $N_{H_{2}}$ $\sim$ 10$^{21}$ cm$^{-2}$ or $A_{\mbox{\scriptsize V}}$ $\sim$ 1 \citep{Andre2010}.
All target fields were mapped in two orthogonal scan directions at a scanning speed of 60$''$ s$^{-1}$ in parallel mode, acquiring data simultaneously in five bands.
The angular resolution in this observing mode is 7.6$''$ at 70 $\mu$m, 11.5$''$ at 160 $\mu$m, 18.2 $''$ at 250 $\mu$m, 25.2$''$ at 350 $\mu$m, and 36.9 $''$ at 500 $\mu$m.
The data were reduced using {\it Herschel} Interactive Processing Environment (HIPE) version 7.0.
A more detailed description of the observations and data reductions is available on the HGBS archives \footnote{ http://gouldbelt-herschel.cea.fr/archives}. 
The reduced SPIRE/PACS maps for all target molecular clouds also are retrieved from the same website.

For the distant and massive clouds, we retrieved the level 2.5 processed, archival {\it Herschel} images, which are from the {\it Herschel} Infrared Galactic plane Survey (Hi-GAL) \citep{Molinari2010PASP}.
The Hi-GAL is a photometric survey designed to map the entire Galactic plane at 70, 160, 250, 350, and 500 $\mu$m simultaneously with 166 individual maps (e.g., ``tiles"), each covering a region of the sky of 2.$^{\circ}$2 $\times$ 2.$^{\circ}$2.
%For the first Hi-GAL public data release (DR1), each tile was processed with the {\tt ROMAGAL} pipeline (Traficante et al. 2011) and photometrically calibrated with the help of IRAS/Planck data 
%The angular resolution is 7.6$''$ at 70 $\mu$m, 11.5$''$ at 160 $\mu$m, 18.2 $''$ at 250 $\mu$m, 25.2$''$ at 350 $\mu$m, and 36.9 $''$ at 500 $\mu$m.
Since we are interested in extended structures, we adopt the extended emission products, which have been absolute zero-point corrected based on the images taken by the {\it Planck} space telescope.

\subsubsection{Calculating \Htwo\ column density with {\it Herschel} images}

We performed single-component, modified black-body spectral energy distribution (SED) fits to each pixel of input {\it Herschel} PACS/SPIRE images.
Before performing any SED fitting, we convolved all images to a common angular resolution of the largest telescope beam (36.9$''$), and all images were re-gridded to have the same pixel size.
We weighted the data points by the measured noise level in the least-squares fits. 
Following \cite{Roy2014}, we adopted the dust opacity per unit mass at 1000 GHz of 0.1 cm$^{2}$ g$^{-2}$ \citep{Ossenkopf1994}.
For the modified black-body assumption, the flux density $S_{\nu}$ at a certain observing frequency $\nu$ is given by
\begin{equation}
S_{\nu} = \Omega_{m}B_{\nu}(T_{\mbox{\scriptsize d}})(1-e^{-\tau_{\nu}}),
\end{equation}
where the column density $N$ can be approximated by
\begin{equation}
N=g(\tau_{\nu}/\kappa_{\nu}\mu m_{H}),
\end{equation}
where $B_{\nu}(T_{\mbox{\scriptsize d}})$ is the Planck function at dust temperature $T_{\mbox{\scriptsize d}}$, the dust opacity $\kappa_{\nu}=\kappa_{\mbox{\tiny{1000 GHz}}}(\nu/1000\,GHz)^{\beta}$, $\Omega_{m}$ is the solid angle, $\mu = 2.8$ is the mean molecular weight, $m_{H}$ is the mass of a hydrogen atom.
We assumed a gas-to-dust mass ratio ($g$) of 100.
The effect of scattering opacity \citep{Liu2019} can be safely ignored in our case given that we are focusing on $>$10$^{3}$ au scale structures, where the averaged maximum grain size is expected to be well below 100 $\mu$m \citep{Wong2016}.

\subsection{Separating bound and unbound gas with N-PDF}\label{subsec:npdf_fit}

The column density probability distribution function (N-PDF) serves as a diagnostic tool to quantify the statistical distribution of gas in molecular clouds \citep{chen2018}.
Both observations and simulations have shown that the N-PDF comprises a low-density lognormal component and one or more high-density power-law components, tracing turbulence-dominated and gravity-dominated gases, respectively \citep{kainulainen2013,lombardi2015,schneider2015}.
This approach effectively separates the bound gas from the unbound gas, and it is the approach we adopt throughout this paper to define and calculate bound gas.
Consequently, the transition point in the N-PDF from a lognormal to a power-law form is considered the column density threshold for star formation. 
Above this threshold, gravity predominates, rendering the gas more likely to collapse and form stars \citep{burkhart2019}.

For the target molecular clouds, we obtained the column density maps with an angular resolution of 36.9$''$, matched to the longest wavelength {\it Herschel} band.
To describe the N-PDF, we used the notation $\eta$, following the frequently used formalism from previous works.
The normalization of the probability function is given by
\begin{equation}
\int\limits_{-\infty}^{+\infty} p(\eta)d\eta = \int\limits_{0}^{+\infty} p(N_{H_{2}})dN_{H_{2}} = 1,
\end{equation}
where the natural logarithm of the ratio of column density and mean column density is $\eta = ln(N_{H_{2}}/\langle N_{H_{2}} \rangle)$ \citep{schneider2015}.
The N-PDF is composed of a low-density lognormal component and a high-density power-law component.
The two components are continuous at the transitional column density \citep{Burkhart2017}.
The distribution can be written as
\begin{equation}\label{eq:ln+pl}
p_{\eta}(\eta) =\begin{cases}
            M(2\pi\sigma_{\eta}^{2})^{-0.5}e^{-(\eta - \mu)^{2}/(2\sigma_{\eta}^{2})} &(\eta < \eta_{t}) \\
            Mp_{0}e^{-\alpha\eta} &(\eta > \eta_{t})
        \end{cases}  ,
\end{equation}
where $\sigma_{\eta}$ is the dimensionless dispersion of the logarithmic field, $\mu$ is the mean, $\eta_{t}$ is the transitional point, $\alpha$ is the slope of the power-law tail, $p_{0}$ is the amplitude of the N-PDF at the transition point, and M is the normalization/scaling parameter.
A few sources' N-PDFs exhibit only the power-law components within their last closed contour. 
For these, we have adopted the column density at the last closed contour as the lower limit for the transitional column density. 
Furthermore, some sources' N-PDFs do not conform to a piecewise function description.
For these cases, we have incorporated an error margin into the transitional column density to account for the uncertainty in the transitional region.

%%%%%%%%%%%%%%%%% figure 1 %%%%%%%%%%%%%%%%%%%
\begin{figure*}[!ht]
\centering
\begin{tabular}{ p{0.45\linewidth}p{0.45\linewidth} }
\hspace{-0.6cm}\includegraphics[scale=0.6]{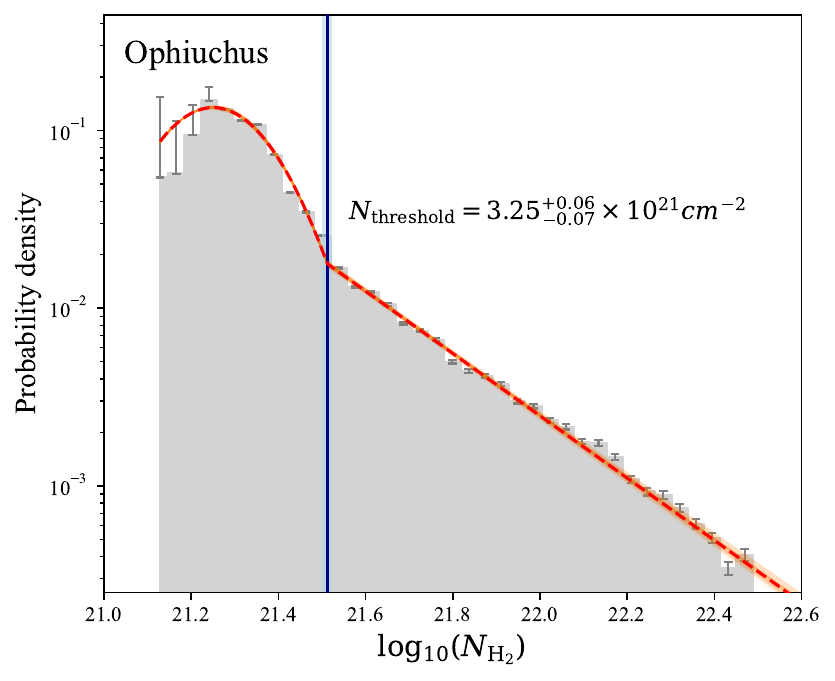} 
& \hspace{-0.1cm} \includegraphics[scale=0.33]{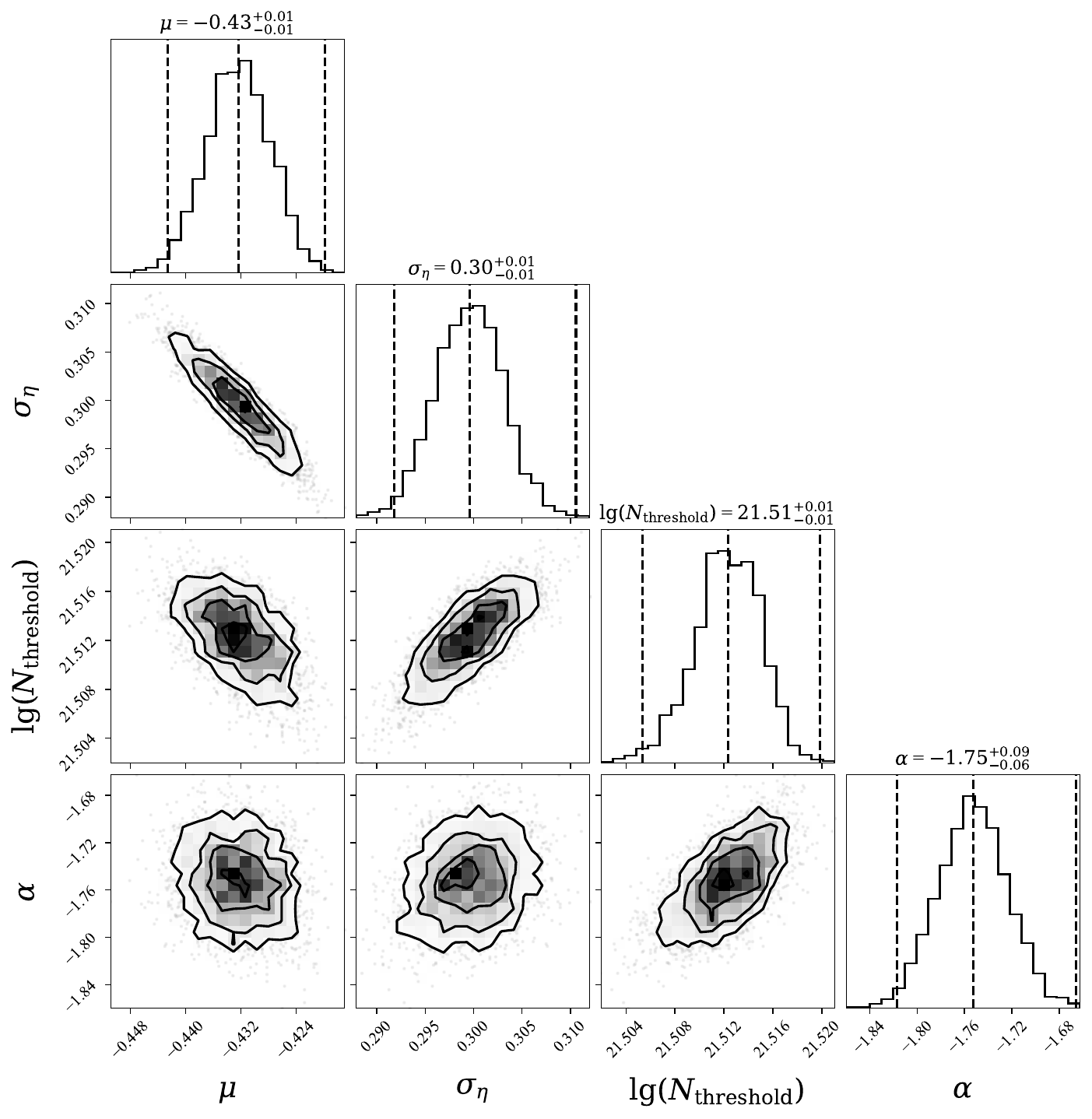}  \\
\end{tabular}
\caption{An example of the N-PDF fitting.
{\it Left:--} The gray histogram shows the N-PDF of Ophiuchus, while the error bars at the lower density end show the uncertainties due to the map area.
The orange line shows the fitted curve and the black vertical line shows the fitted threshold column density.
{\it Right:--}  Corner plot of the fitted parameters.
$\sigma_{\eta}$ represents the dimensionless dispersion, and $\mu$ is the mean of the lognormal component.
$N_{\rm threshold}$ is the transitional column density, related to $\eta_t$ via the relation $\eta_t = \ln(N_{\rm threshold} / \langle N_{\rm H_2} \rangle)$.
$\alpha$ denotes the slope of the power-law tail.
}
\label{fig:mcmc}
\end{figure*}
%%%%%%%%%%%%%%%%% figure 1 %%%%%%%%%%%%%%%%%%%

To account for the sensitivity limits, spatial coverage of the maps, and the requirement for the last closed contour \citep{Alves2017}, we set an optimal column density threshold for each cloud.
We then apply a Maximum Likelihood Estimation (MLE) method \citep{Clauset2009} to fit the N-PDF of each cloud above this threshold.
This approach allows us to fit the N-PDF without pre-binning the data, thus avoiding biases or artifacts that arise from the binning process \citep{Virkar2012ar}.
We employed a Markov Chain Monte Carlo (MCMC) approach to sample the posterior distributions of parameters for this specified model, e.g., Equation \ref{eq:ln+pl}, to derive the transitional column density (both the logarithmic normalized value, $\eta_{t}$, and the absolute value, $N_{\mbox{\scriptsize threshold}}$) and to quantify the uncertainties associated with these estimates.
%Within the Bayesian framework, we employ the Markov Chain Monte Carlo (MCMC) method to sample the posterior distributions of parameters for this specified model, e.g., Equation \ref{eq:ln+pl}.
%This approach allows us to navigate the parameter space to identify the transitional column density (logarithmic normalized transitional column density $\eta_{t}$ and the absolute transitional column density $N_{{\mbox{\scriptsize threshold}}}$) in the N-PDF and to quantify the uncertainties associated with these estimates.
Given the limited prior constraints on the model parameters, we adopt non-informative (uniform) priors for all parameters except for the slope of the power-law tail, $\alpha$.
For $\alpha$, we assume a prior that is uniform in $\tan^{-1}(\alpha)$, which corresponds to a flat prior in the angle of the slope and avoids bias toward steeper values.

The likelihood function is defined as a Gaussian, albeit with a variance that is underestimated by a specific fractional amount $f$.
For each parameter, the median value is taken as the estimate, while the 3-$\mathrm{\sigma}$ error of a Gaussian distribution serves as the parameter's error estimate.
The Python package {\tt EMCEE} \citep{Foreman-Mackey2013} is utilized for the implementation of the fitting process.  

An example of Ophiuchus is shown in Fig. \ref{fig:mcmc}.
Initially, we derived the fitting values using the least squares method. 
Subsequently, we initiated the sampling process by positioning the walkers in a small Gaussian ball centered around the least squares fitting result. 
The primary purpose of this approach is to provide a suitable initial position in the parameter space for MCMC sampling, thereby enhancing the convergence speed and improving the accuracy of posterior distribution estimation. 
To effectively explore the state space of the most critical parameter, $N_{{\mbox{\scriptsize threshold}}}$, particularly because it may exhibit multi-peak distributions, we set the dispersion of the initial position for $N_{{\mbox{\scriptsize threshold}}}$ to be $\sim$ 1000 times larger than that of the other parameters.

We conducted MCMC for each source for 10000 steps, ensuring that the length of the Markov chains is at least 50 times the integrated autocorrelation times for all parameters.
This substantial ratio ensures that the Markov chains have sufficiently converged.
We explored various methods of constructing and updating proposals, referred to as ``moves'' in {\tt EMCEE}, including (a) the classical construction of Metropolis-Hastings proposals that update the walkers using independent proposals, (b) ``stretch move'' ensemble method \citep{Goodman2010CAMCS...5...65G}, (c) a combination of Differential Evolution Move \citep{Nelson2014ApJS..210...11N} and a snooker proposal using differential evolution \citep{terBraak2008}. 
In this work, the third method proved the most effective, enabling the fastest convergence and most effective exploration of the parameter space.
Fig. \ref{fig:lowmass_1} - \ref{fig:highmass_2} display the N-PDFs for the target clouds, which are listed in the Table. \ref{tab:1} and Table. \ref{tab:2}. 

The results of the N-PDF fits are summarized in Table \ref{tab:3}.
Due to differences in cloud distance, the spatial resolution of the {\it Herschel} data varies between the nearby cloud sample and the massive cloud sample.
To evaluate whether this variation in spatial resolution introduces systematic biases in the derived transitional column densities ($N_{\mbox{\scriptsize threshold}}$) and the corresponding $M_{\rm sgb}$ values, we conducted a spatial resolution test using the Ophiuchus cloud as a representative case, as described in Appendix \ref{Appendix:resolution}.

\subsection{Calculating the gravitationally bound gas mass}

The N-PDF can be derived from molecular line emissions, though this method faces challenges due to optical depth and line excitation issues. In contrast, utilizing dust column density measurements offers substantial advantages.
For nearby clouds, the dust column density has been measured by mapping the extinction to background stars \citep{lada2010,Evans2014}. 
However, for more distant clouds, measuring dust column density using the extinction method becomes impractical due to the difficulty of having adequate background stars and separating them from the foreground. 
Consequently, dust emission at far-infrared and submillimeter wavelengths is commonly employed. 
We compared the dense gas masses derived from the N-PDF method using dust emission and from the extinction map, for the sample of nearby star forming regions. These two methods produce consistent results in most nearby clouds.

%%%%%%%%%%%%%%%%% figure 2 %%%%%%%%%%%%%%%%%%%
\begin{figure}
\includegraphics[width=0.97\linewidth]{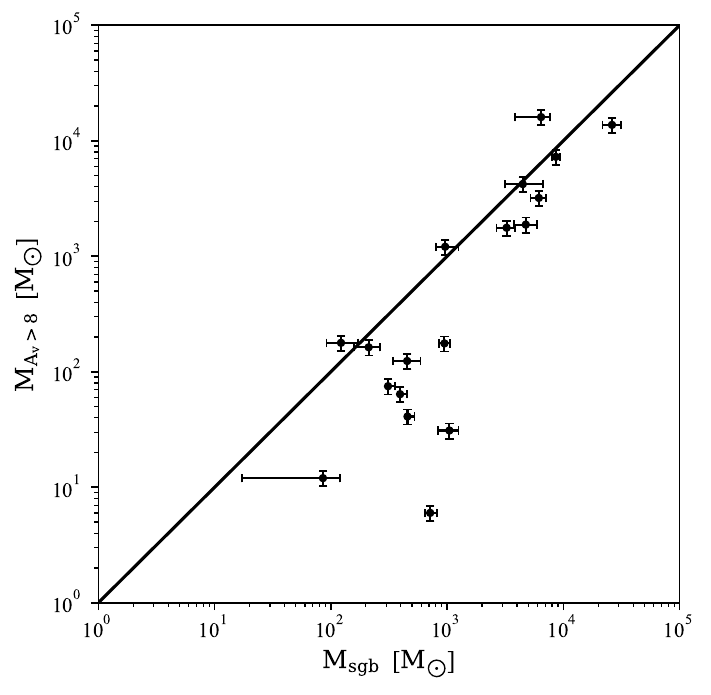}
\caption{Comparison between the derived dense gas mass from ($A_{\mbox{\scriptsize V}}>$8) extinction map \cite{Zucker2021ApJ} and from N-PDF method (this work). 
The black line The black dashed line indicates $M_{\rm A_V>8}$= $M_{\mbox{\scriptsize sgb}}$.
%\textcolor{red}{Label source name if possible?}
}    
\label{fig:N_oph}
\end{figure}
%%%%%%%%%%%%%%%%% figure 2 %%%%%%%%%%%%%%%%%%%

%%%%%%%%%%%%%%%%% figure 2 %%%%%%%%%%%%%%%%%%%
%\begin{figure*}[!ht]
%\begin{tabular}{ p{0.45\linewidth}p{0.45\linewidth} }
%\hspace{-0.3cm}\includegraphics[scale=0.57]{Figures/Av_contour.pdf} & \hspace{0.4cm}\vspace{-0.4cm}\includegraphics[scale=0.72]{Figures/M&M_2025.pdf}  \\
%\end{tabular}
%\caption{
%{\it Left:--} An example of the comparison between $A_{\mbox{\scriptsize V}} = $8 and $N_{{\mbox{\scriptsize threshold}}}$.
%The orange contours represent $A_{\mbox{\scriptsize V}} = $8 , while the blue contours represent the $N_{{\mbox{\scriptsize threshold}}}$.
%{\it Right:--} 
%Comparison between the derived dense gas mass from ($A_{\mbox{\scriptsize V}}>$8) extinction map \cite{Zucker2021ApJ} and from N-PDF method (this work). 
%The black line The black dashed line indicates $M_{\rm A_V>8}$= $M_{\mbox{\scriptsize sgb}}$.
%\textcolor{red}{Baobab: The left and right panels are for very different subjects and should be separated into two figures instead of two panels of one figure. (I understand that they were merged into one figure because Nature and Science do not allow too many figures. Now that we are publishing aa, we should do it in a conventional way.)
%In fact, the left panel of Figure 2 is, to a large extent, duplicated in the top left panel of Figure 3. We should probably simply remove the left panel of Figure 2.
%}
%}
%\label{fig:N_oph}
%\end{figure*}
%%%%%%%%%%%%%%%%% figure 2 %%%%%%%%%%%%%%%%%%%

\begin{figure*}[!ht]
\begin{tabular}{ p{0.45\linewidth}p{0.45\linewidth} }
\includegraphics[scale=0.6]{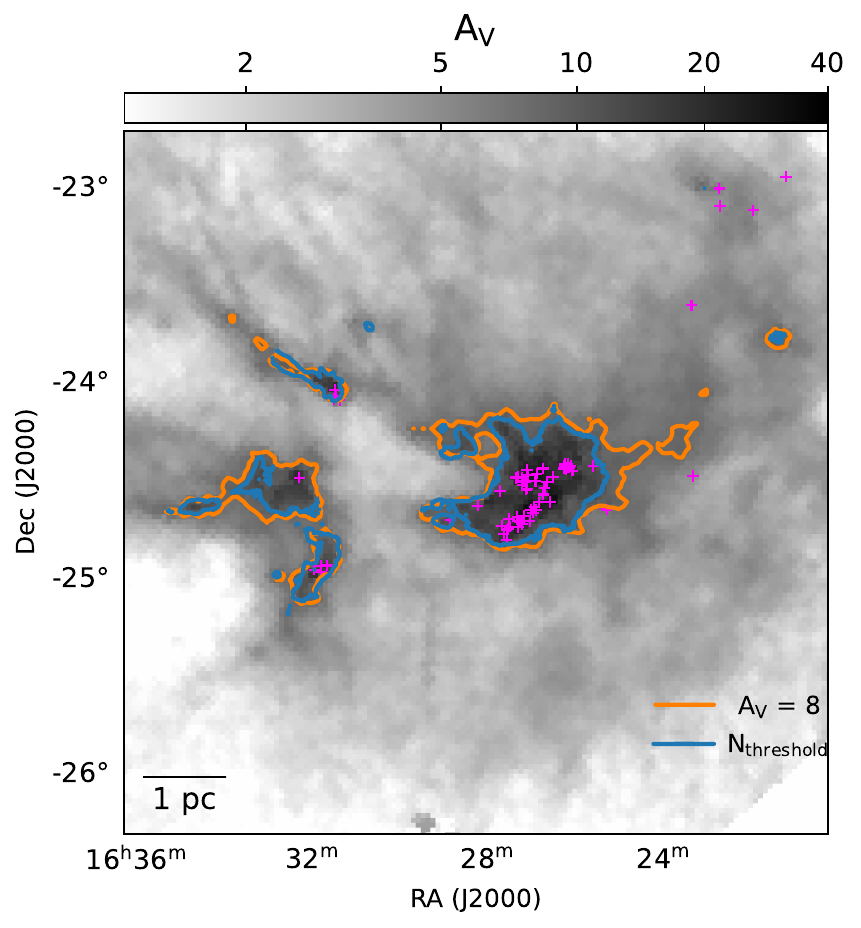} & \includegraphics[scale=0.5]{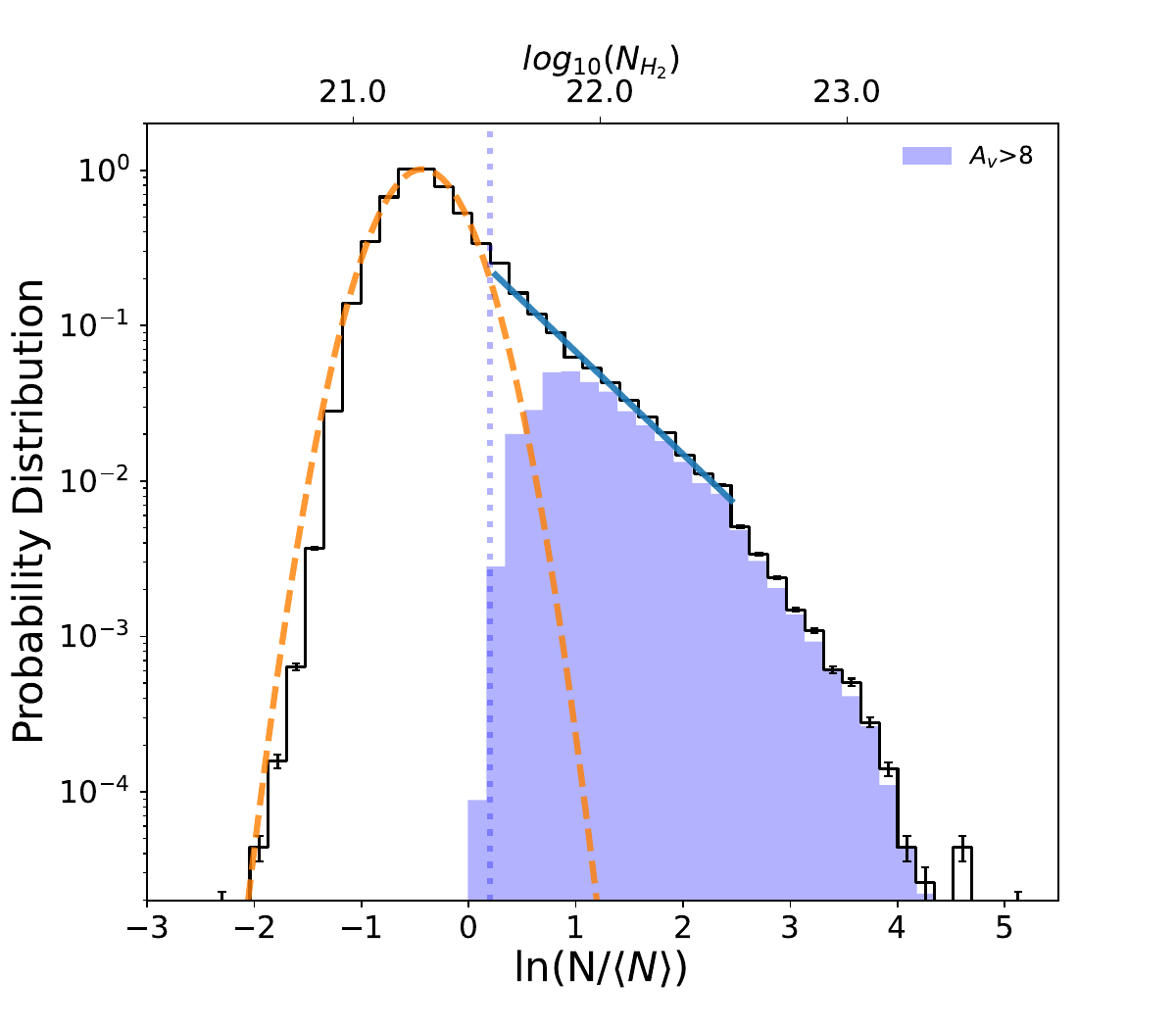}   \\
\end{tabular}
\begin{tabular}{ p{0.45\linewidth}p{0.45\linewidth} }
\hspace{-0.4cm}\includegraphics[scale=0.6]{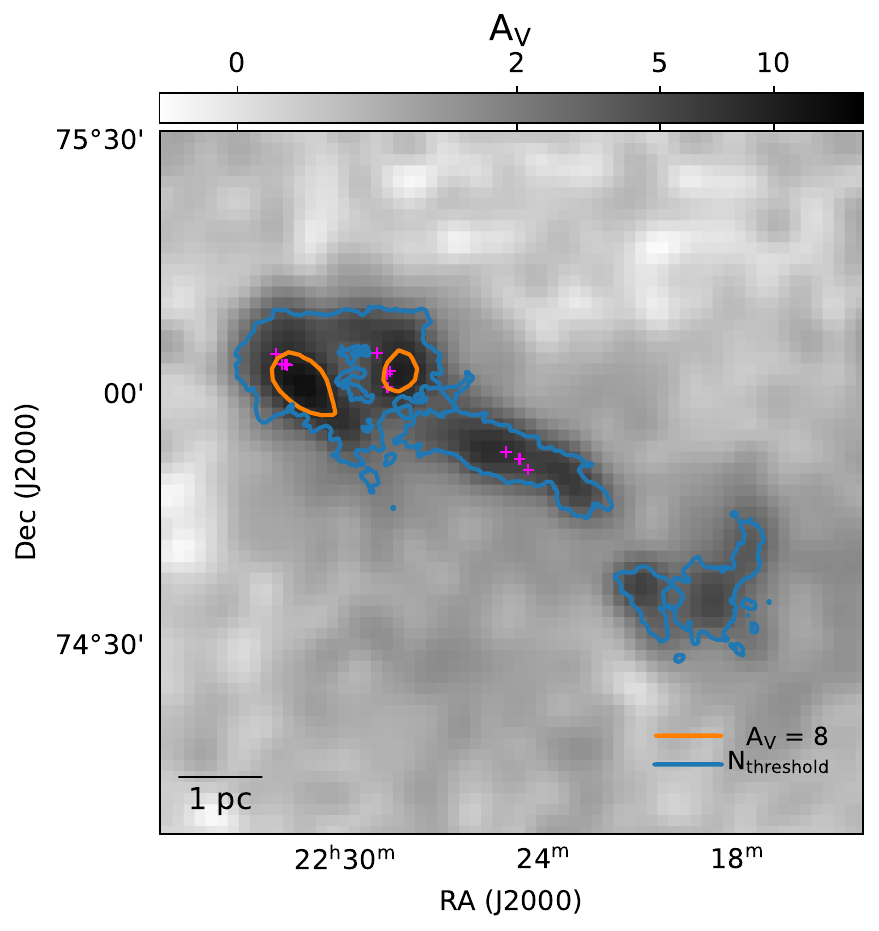} & \includegraphics[scale=0.5]{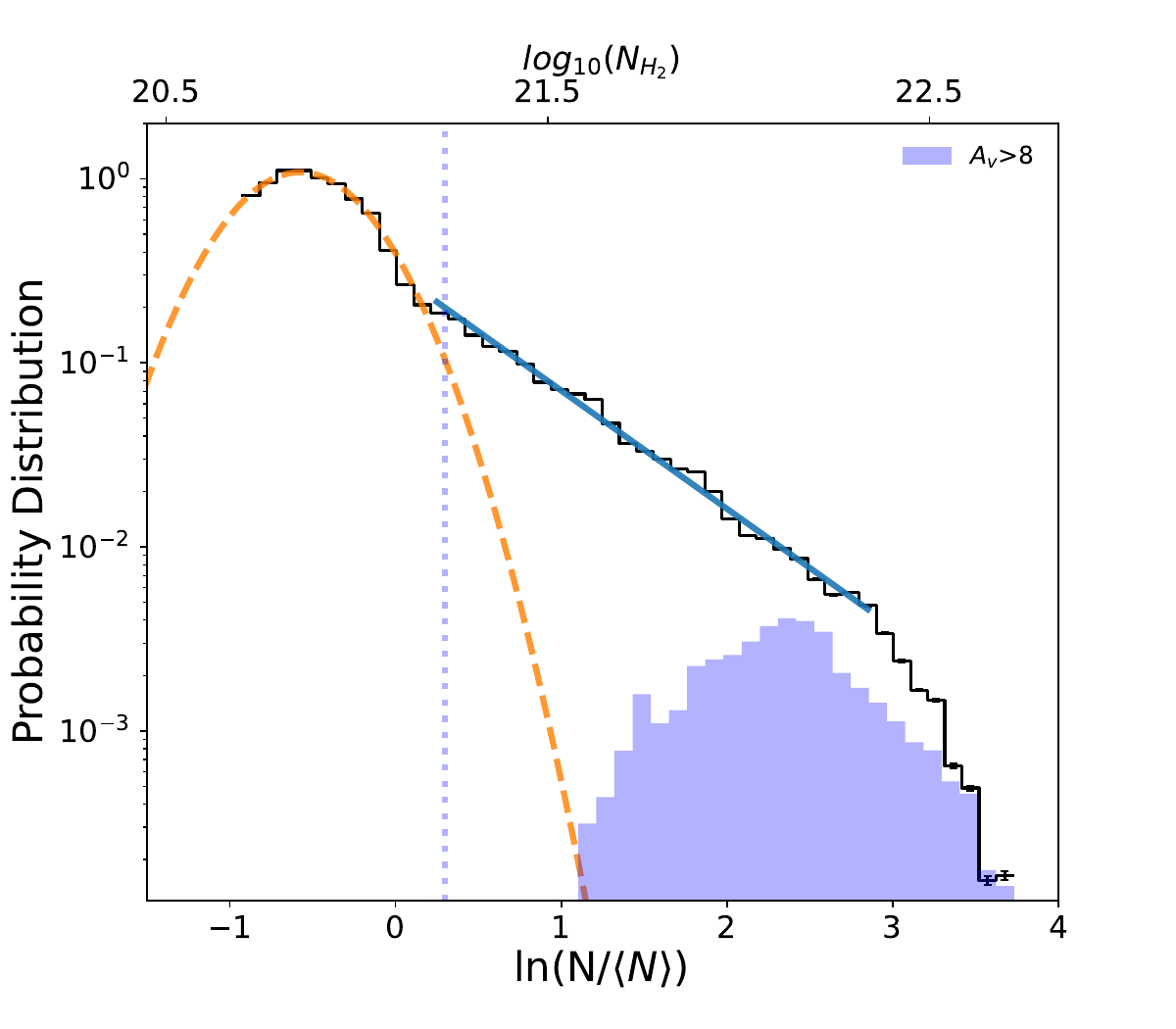}  \\
\end{tabular}
\caption{ 
Example Av maps and N-PDFs.
{\it Left :--} Av maps of Ophiuchus and Cepheus I \citep{Zucker2021ApJ}. 
The orange contours trace $A_{\mbox{\scriptsize V}}$= 8 mag, while the blue contours trace \Nthresh. 
The young embedded protostars (Class 0, I, and Flatspectrum sources) are presented with magenta crosses \citep{Dunham2015ApJS..220...11D}. 
Ophiuchus is a good representative of nearby star-forming clouds, while Cepheus I presents an extreme case with very low SFR.
{\it Right :--} N-PDFs of the Ophiuchus and Cepheus I generated based on {\it Herschel} PACS and SPIRE images (black).  
The fittings of a lognormal component and a power-law tail are demonstrated in an orange dashed line and a blue line.  
We plot the N-PDF of the pixels within the $A_{\mbox{\scriptsize V}}$= 8 mag contour in this panel, which is shown in the grey-blue shade.  
%\textcolor{red}{(1) Left panels need color bars; (2) Adjust the aspect ratios fo the right panels such that they do not look so much shorter than the left panels.}
}
\label{fig:av_PDF}
\end{figure*}

We compared the dense gas structures traced by these two methods based on 2MASS extinction maps and {\it Herschel} column density maps of the clouds in the low-mass star-forming sample.  
An illustrative example is the nearby Ophiuchus, shown in the upper left panel of Fig. \ref{fig:av_PDF}. 
The contours of the extinction-derived $A_{\text{V}}=8$ and dust emission (N-PDF) derived $N_{\text{threshold}}$ coincide closely in the column density map of the cloud. 
This finding is consistent across most nearby clouds in our sample.

%Assuming dust emission is optically thin on cloud scales, 
The mass of self-gravitating gas can be calculated by integrating the mass above the column density threshold (the turning point between log-normal and power-law), i.e., 
\begin{equation}
    M_{\mbox{\scriptsize sgb}} = \int^{+\infty}_{N_{{\mbox{\scriptsize threshold}}}}M(N)dN.
\end{equation}
Fig. \ref{fig:N_oph} displays the comparison between $M_{\rm sgb}^{\rm dust}$ (the mass above $N_{\text{threshold}}$ derived from {\it Herschel} observations) and $M_{\rm A_V>8}^{\rm 2MASS}$ (the mass above an extinction contour of $A_{\mbox{\scriptsize V}}$ = 8 mag, derived from extinction maps and cited from \cite{lada2010} and \cite{Evans2014}). 
Generally, the dense gas mass derived by these two methods aligns well across nearby clouds, except for Cepheus I. 
As also seen in Fig. \ref{fig:SFE}, Cepheus I (the most upper left point) appears as an outlier, exhibiting only a small amount of gas above $A_{\text{V}}=8$ (see also Fig. \ref{fig:av_PDF} ). 
Gas in this cloud becomes bound at a much lower column density than $A_{\text{V}}=8$. 

\section{Results}\label{sec:res}

%%%%%%%%%%%%%%%%% figure 3 %%%%%%%%%%%%%%%%%%%
%%%%%%%%%%%%%%%%% figure 3 %%%%%%%%%%%%%%%%%%%

\subsection{A physical explanation of $A_{\rm V} = 8$ mag threshold}

Fig. \ref{fig:av_PDF} shows the N-PDF analysis in two nearby star-forming clouds, Ophiuchus and Cepheus I. 
A lognormal component and the first power-law component are fit with the breaking point between them, labeled as \Nthresh.   
As the blue-shaded region shows, the break occurs essentially at a column density corresponding to a visual extinction about $A_{\rm V} = 8$  mag for Ophiuchus (as well as for many other low-mass clouds), which is an empirical criterion associated with the best correlation with star formation both in observations for nearby clouds \citep{lada2010,Heiderman2010,Evans2014} and in simulations \citep{Burkhart2017}. 
The fact that $A_{\rm V}= 8$ threshold coincides with \Nthresh\ suggests that this commonly used extinction threshold may not only be an empirical one, but likely has a physical explanation --- above this density, gas becomes bound, and therefore is closely related to star formation.

\subsection{$M_{\rm sgb}$ vs. SFR correlation for Nearby star forming clouds}

For nearby star forming clouds, counting YSOs is so far the most reliable approach to estimate the SFR. We adopted the SFR for the nearby sample from \cite{lada2010}, \cite{Evans2014}, and \cite{Zucker2020}, which used the YSO counting method. 

\begin{figure*}[!ht]
\begin{tabular}{ p{0.4\linewidth}p{0.4\linewidth} }
\includegraphics[scale=0.7]{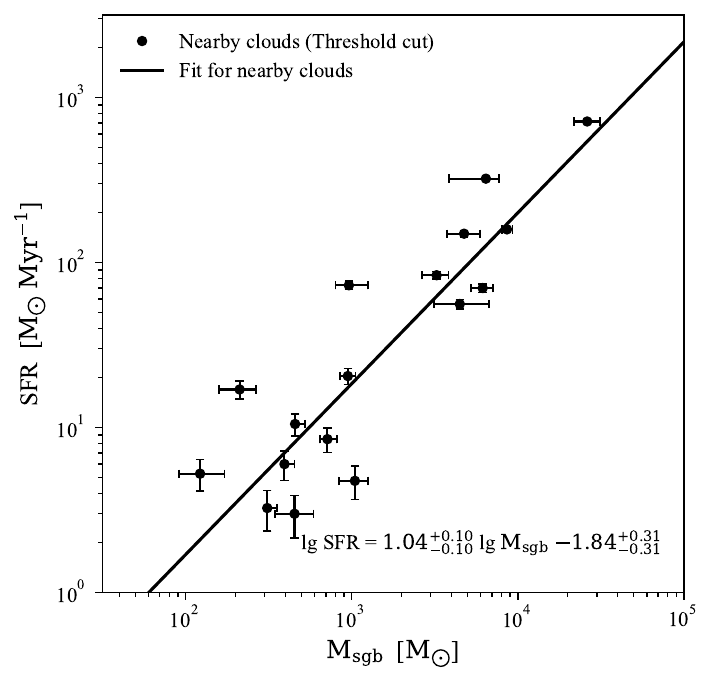} &\hspace{1.cm}\includegraphics[scale=0.7]{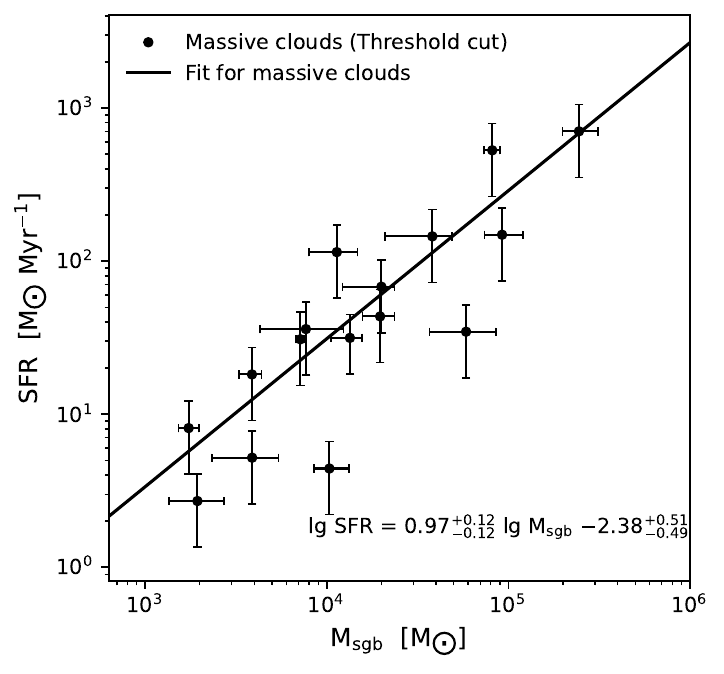}  \\
\end{tabular}
\caption{Correlations between the self-gravitating gas obtained from the N-PDF method, and the star formation rates, for 
{\it Left:--} A sample of nearby massive star-forming regions. 
The black line shows a linear least-squares fit with a slope of 1.04$\pm$0.10.
{\it Right:--} A sample of massive star-forming regions. 
The black line shows a linear least-squares fit with a slope of 0.97$\pm$0.12.
%\textcolor{red}{Label source name if possible? Like what was done in the MMC paper?}
}
\label{fig:mass_sfr}
\end{figure*}

We apply a linear least-squares fit to the correlations between the self-gravitating gas obtained from the N-PDF method ($M_{\mbox{\scriptsize sgb}}$) and SFR (using YSO counting) for the low-mass clouds in our sample, resulting in a tight, linear correlation with a slope of 1.02$\pm$0.10 (see the left panel of Fig. \ref{fig:mass_sfr}). 
The fitting process accounts for errors associated with each measurement. 

%%%%%%%%%%%%%%%%% figure 4 %%%%%%%%%%%%%%%%%%%
%%%%%%%%%%%%%%%%% figure 4 %%%%%%%%%%%%%%%%%%%

This result is consistent with previous works using the extinction map and a fixed \av = 8 mag cut to calculate dense gas mass, and the YSO counting method for SFR, for a sample of nearby star forming clouds. 
They also obtained a linear correlation, with slopes of 0.96$\pm$0.13 \citep{lada2010} and 0.89 \citep{Evans2014}. 
However, although \Nthresh\ in our nearby sample roughly meet the $A_{\rm V}> 8$ mag criterion, there is still substantial variation ranging from 1.0 to 11.0 $\times$ 10$^{21}$ cm$^{-2}$ in column density (See Table \ref{tab:1}), consistent with \cite{Schneider2013} which noticed the variation for a much smaller sample. 
The mean and median of \Nthresh\ for the nearby clouds are 3.50 $\times 10^{21} cm^{-2}$ and 2.05 $\times 10^{21} cm^{-2}$, respectively.
For example, in the case of Cepheus I as presented in Fig. \ref{fig:av_PDF}, the \Nthresh\ is much smaller than  $A_{\rm V}= 8$ in this extremely low SFR cloud, but we can still see star formation occurring in regions with $A_{\rm V} < 8$ mag but  N >\Nthresh\, indicating bound gas traces star-forming gas better than $A_{\rm V}= 8$ in this cloud. 
The variation of \Nthresh\ in nearby clouds likely reflects the influence of turbulence within these regions. 
This urges us to extend the dynamic range of dense gas mass and SFR to test the correlation in high-mass star-forming clouds.

\subsection{$M_{\rm sgb}$ vs. SFR correlation for high-mass star forming clouds}

For distant and massive star forming regions, no correlations between the extinction-based dense gas mass and SFR have been investigated before, because both extinction map and counting YSOs to compute the SFR are unfeasible for distant clouds. We rely on bolometric luminosity to estimate the SFR for massive clouds. The idea is that the luminosity of stars is absorbed by dust in the molecular cloud and re-radiated in the far-infrared \citep{KE2012}. 
For a sample of massive star forming regions \citep{Wu2010}, we derive SFRs from their bolometric luminosity by using the relation SFR $\approx$ 2 $\times$ 10$^{-10}$($L_{\mbox{\scriptsize IR}}/L_{\odot}$) $M_{\odot}$ yr$^{-1}$ from IRAS flux, following \cite{Gao2004a} and \cite{Kennicutt1998}. 
We also use the N-PDF method to separate unbound and bound gas, to derive their bound gas mass. 
Uncertainties of distance and dust contamination from fore- and background sources only contribute to minor errors, and they do not change the conclusions (See Appendix for details). 
The uncertainty in SFR(L$_{\rm bol}$) arises from several factors: measurement errors in flux, uncertainties in distance, and the uncertainty of the conversion from the bolometric luminosity to SFR.
Due to the lack of reliable uncertainty of the conversion from the bolometric luminosity to SFR, we have assumed a 50\% uncertainty for each high-mass, distant cloud in our SFR measurements.

We also made a linear least-squares fit of $M_{\mbox{\scriptsize sgb}}$-SFR correlation to the high-mass clouds in our sample, with $M_{\mbox{\scriptsize sgb}}$ derived from the N-PDF method and SFR calculated from bolometric luminosity. 
The fitting result is also linear, with a slope of 0.97$\pm$0.11 (Fig. \ref{fig:mass_sfr}). 

Although we apply different methods to derive SFR in nearby clouds and in distant, massive clouds, we see a clear trend that bound gas mass has a tight, linear correlation with SFR, presenting consistent star formation efficiency (SFR per amount of bound gas) no matter which method we use. Our results demonstrate that bound gas mass is closely related to star formation, for both nearby, low-mass clouds and distant, high-mass clouds.

\section{Discussion}\label{sec:dis}

\subsection{ Comparing SFR for low-mass and high-mass star-forming regions}\label{sub:SF}

\begin{figure}
\includegraphics[width=0.97\linewidth]{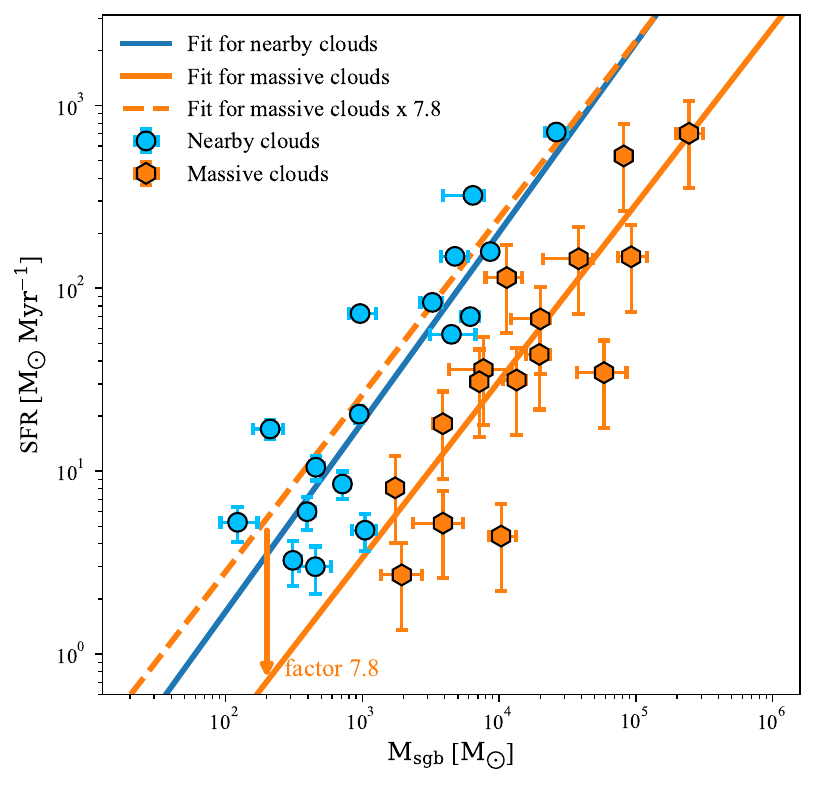}
\caption{Correlations between the self-gravitating gas obtained from the N-PDF method, and the star formation rates, for both nearby clouds using the YSO counting method, and distant, massive clouds using the bolometric luminosity method.
Both correlations are almost linear, with an offset of $\sim$7.8.
}    
\label{fig:SFrelaltion}
\end{figure}

For both nearby clouds and distant, massive clouds, mass of the gas belongs to the power-law component of the N-PDF has a tight, linear correlation with SFR. 
In Fig. \ref{fig:SFrelaltion}, we present these two linear correlations together. 
Apparently, there is an offset of a factor of 7.8 in SFR between them.  
As we show in Appendix \ref{Appendix:resolution}, the measurement of \st{bound} gas mass in the two cloud samples is consistent, so that the different intercepts are unlikely due to the resolution of imaging. 
More likely, this offset reflects the difference in measuring SFR in these two systems.

%\textcolor{red}{I think we should omit this paragraph and absorb some useful references in Section 4.2. First, we do not know the ages of the high-mass sample. They can be younger than some or most of the solar neighborhood star-forming regions. 
%We do not know the stellar population in the high-mass sample either.
%Finally, too many statements are in relative instead of absolute sense (e.g., young vs. old, fully-sampled vs. incomplete, etc, which do no help understand the systematics). 
%My suggested revision is embedded in Section 4.2.
%}
 
Direct comparison between SFR (YSO) and SFR (IR) is quite complicated, given differences in time scale and sensitivity to the IMF of the two methods. 
Massive star formation tracers, such as infrared luminosity, only reflect SFR well for clumps massive enough to have a fully sampled IMF, and old enough (5$-$10 Myr) to reach statistical steady state \citep{Krumholz2007ApJ}, thus normally fails to trace SFR in low-mass clouds \citep{Wu2010, Gutermuth2009ApJS, Gutermuth2011ApJ}. 
In contrast, the YSO counting method for SFR focuses on small scales and shorter time scales (typical for YSOs of a couple of Myr), and trace star formation well in low-mass regions (\cite{lada2010, Heiderman2010, Evans2014}).

%\brevise{
%It has been known that massive  star formation tracers, such as infrared luminosity, only reflect SFR well for clumps massive enough and normally fails to trace SFR in low-mass clouds \citep{Krumholz2007ApJ,Wu2010, Gutermuth2009ApJS, Gutermuth2011ApJ}. 
%}

Significant difference between the dense gas-SFR correlations based on massive star formation SFR tracers (like infrared luminosity) and smaller-scale SFR tracers has also been reported by other works \citep[e.g.,][]{lada2012, Pokhrel2021ApJ, Elia2025}. 
For example, \cite{lada2012} found that the linear correlation between the SFR(YSO) and dust-extinction derived cloud mass for nearby clouds and the Gao-Solomon correlation between SFR(IR) and dense gas (from HCN) for galaxies \cite{Gao2004a} has an offset of 2.7. 
Given that both linear relations span a large range of magnitudes in mass, with coefficients being consistent within quoted errors, \cite{lada2012} argued that they should represent the same relation. Similar to Lada’s arguments, we believe the nearby clouds and distant, massive clouds in our sample follow similar underlying physical processes, given that the two samples have overlaps in bound gas mass range, that the two linear correlations likely reflect the same relation. 
Especially that for the only sources in our sample (Orion A and Orion B) that happen to have been measured in both methods, \cite{lada2012} found the SFRs measured in the two methods differ by a factor of $\sim$8, which is quite consistent with the discovered offset of 7.8. 
Yet we should keep in mind that such a constant conversion from SFR(YSO) to SFR(LIR) has not been fully justified given very different spatial and time scales, and has to be used in caution.

%\textcolor{red}{This paragraph makes the discussion about SFE strange in Section 4.2 strange. I suggest remove it and absorb the necessary discussion in Section 4.2. The last sentence is not specific and is over speculative, which should only be in the reply to referee.}
%However, we do not have sufficient evidence yet to claim a constant conversion factor from SFR(YSO) to SFR(LIR), given that they deal with very different spatial and time scales. 
%The YSO-counting based SFR has variations that depend on different strategies to count young stars \citep[e.g.,][]{Pokhrel2020ApJ}.
%The major origin for the offset of 7.8 between correlations may come from an averaging effect, both on the time scale and spatial scales over the entire cloud, awaiting further investigations.
%Although we can't quantitatively explain the factor of 7.8 difference in the two correlations, it doesn't affect our conclusion that bound gas nicely traces star formation rate, both on small scales and large scales.

%Comparing SFR-related correlations between different systems where SFRs are calculated from different methods (YSO counting and bolometric luminosity methods for example) requires a conversion of SFR measurements between these two systems, which has not been fully established. 

%%%%%%%%%%%%%%%%% figure 5 %%%%%%%%%%%%%%%%%%%
%%%%%%%%%%%%%%%%% figure 5 %%%%%%%%%%%%%%%%%%%

%The tight, linear correlations between the gravitationally bound gas mass and the SFR for both nearby, mostly low-mass clouds and distant, massive clouds suggest that the star formation efficiency per amount of bound gas is consistent in these two systems, despite the different methods to calculate their SFRs. 
%If we believe that star formation in nearby clouds and distant clouds follow the same physical process, given that both fits are linear that indicating a constant efficiency to convert bound gas into stars in their own systems, the offset of the two linear fittings in intercepts likely reflects the systematic shift in calculating SFR with the two methods.
%In order to keep the \Msgb-SFR correlation to be linear within both systems, we should only apply a constant conversion factor between these two SFRs, which can be obtained from the difference in the intercepts of the two fittings. 
%This gives a conversion factor of 7.8$\pm$0.7, which is SFR(YSOs) = 7.8 $\times$ SFR(L$_{\rm bol}$). 

%%%%%%%%%%%%%%%%% figure 6 %%%%%%%%%%%%%%%%%%%
\begin{figure*}[!ht]
\centering
\includegraphics[width=0.95\linewidth]{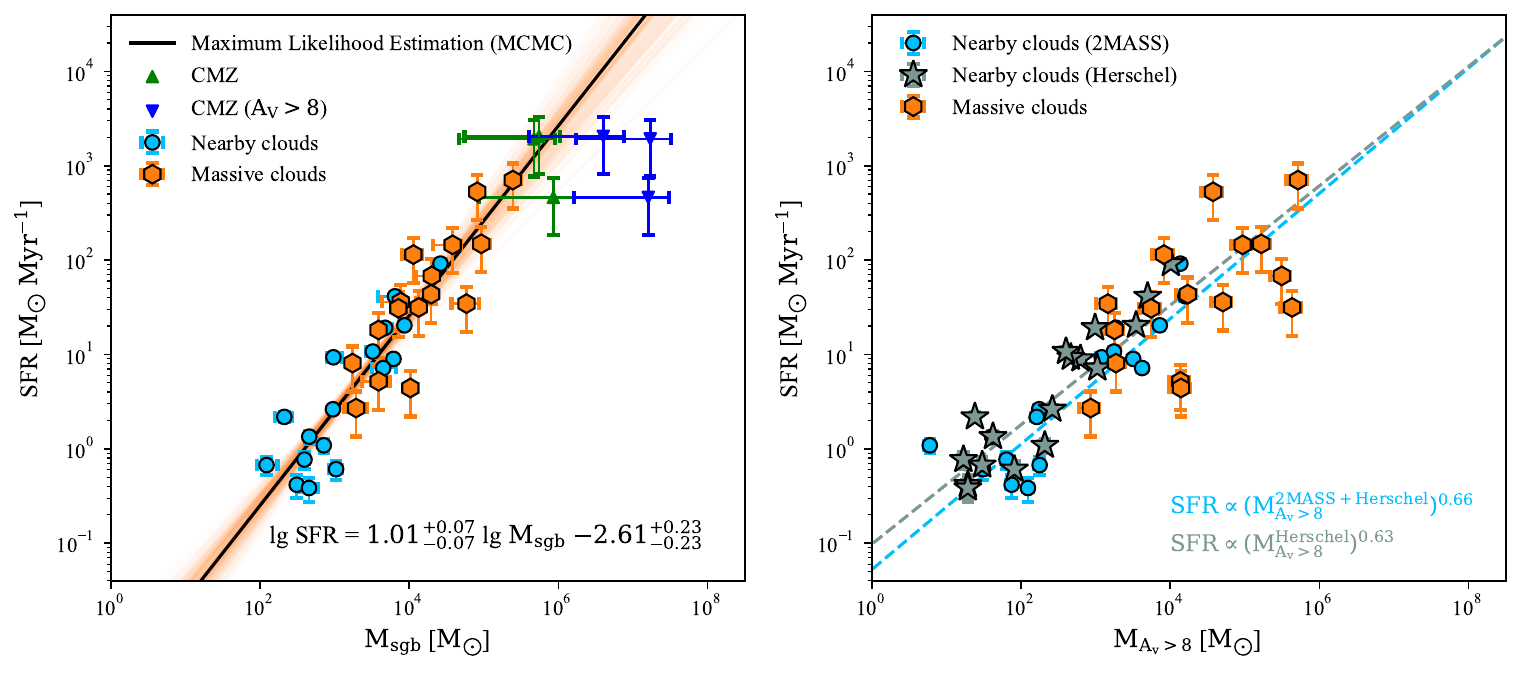}
\caption{
SFR vs. dense gas mass correlations.
{\it Left:--} The correlation between the self-gravitating gas mass obtained from N-PDF and the star formation rates, for both nearby and high-mass star-forming regions.  
The black line is a fitting for both nearby and massive clouds, with a slope of $1.01\pm0.07$.  
Three regions from CMZ are marked in the plot, whose SFRs from the literature \citep{Longmore2013}, and dense gas mass calculated by the N-PDF method (green triangles) and by a fixed threshold method (blue triangles).
{\it right:--} Similar to the left panel, but the dense gas masses are calculated by a fixed extinction or column density threshold method. 
{\bf The dashed lines show the fitting with a slope = 0.66 for $M_{\rm A_V>8}^{\rm 2MASS+Herschel}$ and 0.63 for $M_{\rm A_V>8}^{\rm Herschel}$.}
%\textcolor{red}{In the left paanel, label the names of the CMZ sources?}
}
\label{fig:mass_sfr_all}
\end{figure*}
%%%%%%%%%%%%%%%%% figure 6 %%%%%%%%%%%%%%%%%%%

%In the only massive star forming clouds in our sample, Orion A and B, whose SFR(YSOs) and SFR(L$_{\rm bol}$) have both been measured, the ratio of SFR(YSOs) to SFR(L$_{\rm bol}$) is $\sim$8.0 \citep{lada2012}
%This aligns closely with our findings, supporting our conversion between these two SFR systems. 
%Consequently, we apply this conversion factor of 7.8 to the SFRs of massive clouds and list the corrected SFRs in Table \ref{tab:2}. 

\subsection{A consistent star formation efficiency in terms of bound gas in Galactic molecular clouds}

%\textcolor{red}{This repeats the information in the previous section.}
%\st{
%For both samples of nearby, mostly low-mass and distant, massive star forming clouds, the bound gas has a nice linear correlation with SFR, suggesting that star formation has a consistent efficiency to form stars in terms of bound gas, in both nearby and distant massive clouds. 
%}

%\textcolor{red}{I suggest removing this paragraph. I merge some useful information into the next paragraph. In general, we should first state what we tried, then what we found, and an interpretation/explanation for what we found. Starting with explanations reads like excuses, which do not give a good impression.}
%{\bf 
%For convenience to compare star formation rate in the two samples, especially to reflect the consistent efficiency to convert bound gas into stars, we match the offset between the two linear fits for nearby clouds and massive, distant clouds in Fig. \ref{fig:mass_sfr_all}, by converting SFR in YSO counting system into IR-based SFR system by dividing the factor 7.8. 
%The reason to convert YSO system into massive star formation system is because the large-scale, averaged SFR is more interesting to be compared to SFR over the GMC scale or in galaxies. Yet we need to keep in mind that such a constant conversion of SFRs between the two systems has not been fully justified and should be used with caution, which needs more detailed studies in the future. 
%}

%\brevise{

If we naively scaled down the SFR(YSO) of the solar neighborhood sample by a factor of 7.8, to match the SFR(IR) for massive clouds and that most extragalactic study used, the SFR(YSO) and SFR(IR) of Orion A and B are also matched.
We are intrigued to see a tight linear correlation in the plot of \Msgb - SFR (Fig. \ref{fig:mass_sfr_all}) formed by the 36 star-forming clouds (from low-mass to high-mass regions over four orders of magnitude), in spite that there is no evidence that the SFR(YSO) in all neighborhood clouds can be correctly re-scaled to SFR(IR) by multiplying a constant factor.
A linear regression yielded a slope of $0.97\pm0.07$.
Based on this result, and the fact that the \Msgb - SFR correlation is linear and tight for both the solar neighborhood sample and high-mass sample, we argue that the power-law component in the N-PDFs may define the star-forming gas in a molecular cloud, which is likely gravitationally bound.

% After converting their SFRs into the same system, we {\bf try to} combine the nearby and distant samples with a single fitting for \Msgb - SFR , assuming they are internally following the same correlation. 
% The fitting for the 36 star-forming clouds (from low-mass to high-mass regions over five orders of magnitude) shows a {\bf very} tight, linear correlation (Fig. \ref{fig:mass_sfr_all}), with a slope of $0.97\pm0.07$. 
% We define the N-PDF obtained, gravitationally dominated bound gas as star-forming gas, in contrast to the traditionally defined dense gas with a fixed extinction or column density threshold.  
% Fig. \ref{fig:mass_sfr_all} shows the $M_{\mbox{\scriptsize dense}}$-SFR correlation for the N-PDF derived star-forming gas and the dense gas with a fixed threshold. 

\begin{figure*}[!ht]
\centering
\includegraphics[width=0.99\linewidth]{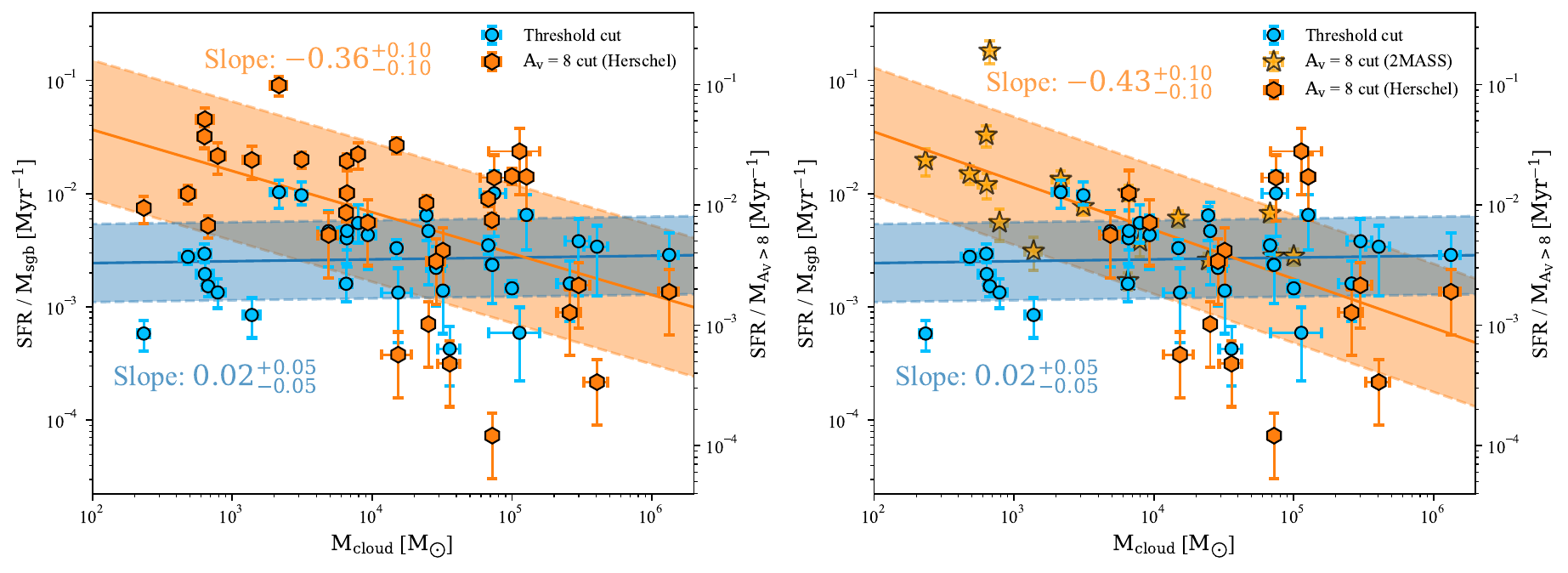}
\caption{
Star formation rate per unit dense gas mass for bound gas and gas with a fixed density threshold.
{\it Left:--} $M_{\rm A_V>8}$ derived from {\it Herschel} column density maps above $N(\rm H_{2})$ = 7.4$\times$10$^{21}$ cm$^{-2}$.
The cloud mass M$_{\mbox{\scriptsize cloud}}$ is calculated by integrating all cloud mass above $A_{\mbox{\scriptsize V}}$ = 1. 
Blue circles show the relationship based on \Msgb from the N-PDF method, and the orange hexagons show the relationship based on $M_{\rm A_V>8}^{\rm Herschel}$.  
The orange and blue shaded regions correspond to the regimes with 1 $\sigma$ data scatters in linear least-squares fittings.
{\it Right:--} Same as the left panel but for $M_{\rm A_V>8}^{\rm 2MASS+Herschel}$ case.
}
\label{fig:SFE}
\end{figure*}

To conduct a comparison with literature results derived based on a fixed threshold, we adopted the $A_{\mbox{\scriptsize V}}$ = 8 mag cut to estimate the dense gas mass ($M_{\rm A_V>8}$). 
For nearby clouds, both $A_{\mbox{\scriptsize V}}$ and {\it Herschel} column density maps are available.
We used the dense gas masses enclosed within the the extinction contour of $A_{\mbox{\scriptsize V}}$ = 8 mag from the 2MASS extinction maps, as reported by \cite{lada2010} and \cite{Evans2014}, and refer to these as $M_{\rm A_V>8}^{\rm 2MASS}$.
Using the standard conversion of $N(\rm H_{2})/A_{\mbox{\scriptsize V}}$ = 0.94$\times$10$^{21}$ cm$^{-2}$ mag$^{-1}$ \citep{Frerking1982}, we applied a column density threshold of 7.4$\times$10$^{21}$ cm$^{-2}$ to the {\it Herschel} maps to obtain $M_{\rm A_V>8}^{\rm Herschel}$.
%We take a standard conversion between $A_{\mbox{\scriptsize V}}$ and column density $N(\rm H_{2})/A_{\mbox{\scriptsize V}}$ = 0.94$\times$10$^{21}$ cm$^{-2}$ mag$^{-1}$ \citep{Frerking1982}, thus cut the dense gas mass above the threshold at 7.4$\times$10$^{21}$ cm$^{-2}$ on the {\it Herschel} column density maps as $M_{\rm A_V>8}^{\rm Herschel}$.
For massive clouds, where $A_{\mbox{\scriptsize V}}$ maps are not available, we adopted the same threshold directly on the {\it Herschel} column density maps to derive $M_{\rm A_V>8}^{\rm Herschel}$.
We found that the correlation between SFR and the bound gas mass (\Msgb) is linear and significantly tighter than the correlation using the fixed-threshold dense gas mass ($M_{\rm A_V>8}$).
%The fitted correlation between $M_{\mbox{\scriptsize dense}}$ and SFR is a linear correlation and is much tighter for bound gas mass (\Msgb) than the correlation for dense gas mass from a fixed threshold ($M_{\rm A_V>8}$).  
The latter correlation is non-linear, with a slope = 0.66 for $M_{\rm A_V>8}^{\rm 2MASS+Herschel}$ ({$M_{\rm A_V>8}$ derived from extinction maps above $A_{\mbox{\scriptsize V}}$ = 8 mag for nearby clouds and from \it Herschel} column density maps above $N(\rm H_{2})$ = 7.4$\times$10$^{21}$ cm$^{-2}$ for massive clouds) and 0.63 for  $M_{\rm A_V>8}^{\rm Herschel}$ ({$M_{\rm A_V>8}$ derived from \it Herschel} column density maps above $N(\rm H_{2})$ = 7.4$\times$10$^{21}$ cm$^{-2}$ for all clouds), and exhibits a larger scatter. 
%This is likely due to the larger contribution and variation of turbulence in massive star forming regions, that only a fraction of dense gas actually contributes to star formation. 
This difference is more clearly illustrated by comparing the star formation efficiencies (i.e., SFR per unit dense gas mass), as shown in Figure \ref{fig:SFE}.
%The difference is better seen when comparing their star formation efficiencies (SFR per amount of dense gas mass), as presented in Fig. \ref{fig:SFE}. 

%%%%%%%%%%%%%%%%% figure 7 %%%%%%%%%%%%%%%%%%%
%%%%%%%%%%%%%%%%% figure 7 %%%%%%%%%%%%%%%%%%%

%\textcolor{red}{Some information is repeated in the next paragraph, which I crossed out.}
%\st{
%The dense gas star formation efficiency is almost constant for the bound gas, but is nonlinear with a much larger scatter for dense gas obtained with a fixed column density, suggesting that the amount of turbulence varies substantially.} 
 Our tentative hypothesis is that for some solar neighborhood clouds, using a fixed column density threshold of $A_{\mbox{\scriptsize V}}>$ 8 mag led to underestimates of the amount of star-forming gas (see Cepheus I in Fig. \ref{fig:av_PDF} as an example).
This might because these clouds are less turbulent such that molecular gas can be bound by self-gravity at considerably lower densities than what is inferred by the $A_{\mbox{\scriptsize V}}=$ 8 mag criterion.
On the other hand, massive star-forming molecular clouds may be more turbulent such that only gas in higher density regions can be bound by self-gravity.
In such cases, sticking to a fixed column density will overestimate the amount of star-forming gas.
The most extreme cases may be those in the central molecular zone (CMZ), which are discussed in Section \ref{sub:CMZ}.

% Towards the higher-mass end, the threshold of bound gas may increase due to the more turbulent environment, so sticking to a fixed column density will overestimate the amount of star-forming gas.

%\textcolor{red}{First sentence is an unnecessary repeat of the hypothesis introduced in Section 1.}
%\st{Using the N-PDF, we have separated the mass of the molecular clouds into bound (self-gravitating) and unbound portions.}
%\brevise{

We based on Fig. \ref{fig:mass_sfr_all} to derive the star-forming efficiency (SFE, i.e., SFR$/M_{\rm sgb}$) in the bound gas traced by the power-law component of the N-PDF.
We found that across a large range of cloud parameters and Galactic environments, the SFE is nearly constant.
The derived mean and median SFR are both 0.004 Myr$^{-1}$, meaning that {\bf $\sim$0.4\%} of the enclosed gas mass is converted to stars every megayear, although this derivation is up to the uncertainties in converting YSO counts and infrared luminosities into SFR (c.f. see discussion in Section \ref{sub:SF}).

%}
%\st{
%Based on Fig. \ref{fig:SFE}, we can see that over a large range of cloud parameters and Galactic environments, the bound gas \brevise{traced by the power-law component of the N-PDF} form stars at a nearly constant efficiency, with the {\bf median value of 0.004 Myr$^{-1}$ and average 0.004 Myr$^{-1}$}.
%This implies that the bound gas converts {\bf $\sim$0.4\%} of the total content into stars per million years, {\bf based on the large-scale SFR. 
%If we only consider the small scale, short time scaled SFR, the corresponding mean and median star formation efficiency would be 0.03 Myr$^{-1}$ and 0.03 Myr$^{-1}$}.
%}

\subsection{The Central Molecular Zone areas}\label{sub:CMZ}

The N-PDF method can naturally explain the very low star formation efficiency in the central molecular zone close to the Galactic center.  
\cite{Longmore2013} studied three large areas in the CMZ, finding that their star formation rate is much lower than predicted by their dense gas content.
The SFR is an order of magnitude lower than predicted for gas with $A_{\mbox{\scriptsize V}}>$8 mag. 

\cite{Longmore2013} argued that the low SFR is because the CMZ is very turbulent, and star formation can only occur in very dense regions where gravity can overcome the high turbulence.
This idea has been widely explored \citep[e.g.,][]{Liu2013ApJ...770...44L,Rathborne2014ApJ,Henshaw2016MNRAS,Federrath2017IAUS,Walker2018MNRAS,Barnes2019MNRAS,Lu2020ApJ,Henshaw2023ASPC}.  
In particular, \cite{Rathborne2014ApJ} used high-sensitivity ALMA 3 mm continuum observations to study the N-PDF of G0.253+0.016, an exceptionally massive and dense molecular cloud in the CMZ that shows no clear evidence of widespread star formation \citep{Walker2021MNRAS}.
They found that there is a small deviation from the log-normal distribution at the highest column densities, indicating self-gravitating gas, which exactly coincides with the location of the H$_2$O maser emission \citep{Lu2019ApJ}.
Using the N-PDF method, we extend the analysis from CMZ to our sample and find the threshold column density for the $-$1$^{\circ}$ $< \it{l} < $ 1$^{\circ}$, $\vert b \vert$ $<$ 0.5$^{\circ}$ area is about 40 times higher than that of Orion B.

Similar concerns arise regarding the consistency of SFR calculations in the CMZ regions and our other samples.  In the three CMZ regions, SFRs are measured through radio free-free emission using Wilkinson Microwave Anisotropy Probe (WMAP) Sky Maps, so we need to quantify the potential biases in the SFRs measured using the radio free-free method, and the YSO counting, bolometric luminosity approaches.

%%%%%%%%%%%%%%%%% figure 8 %%%%%%%%%%%%%%%%%%%
\begin{figure}
\includegraphics[width=0.98\linewidth]{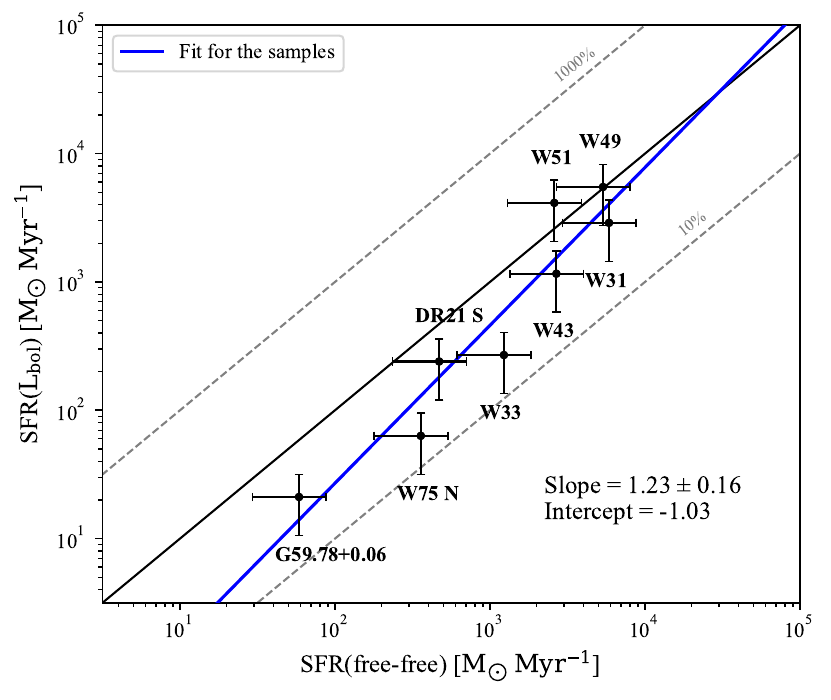}
\caption{Comparison between SFR(L$_{\mbox{\scriptsize bol}}$) and SFR(free-free).
The solid black line represents the 1–1 line where the two SFRs are equal. 
The two grey lines represent 10 and 0.1 times the SFRs from the infrared and free-free emissions, respectively. 
The blue line is a linear fit to the data.
}    
\label{fig:radio_ir}
\end{figure}
%%%%%%%%%%%%%%%%% figure 8 %%%%%%%%%%%%%%%%%%%

Given the relatively low resolution of WMAP, our selection is limited to isolated sources with considerable angular sizes. We identified six massive star-forming clouds with available WMAP W-band measurements and estimated their SFRs following the approach outlined by \cite{Rahman2010}.
Fig. \ref{fig:radio_ir} presents a comparison between the SFRs derived from free-free emission and bolometric luminosity for these massive clouds. 
SFRs from these methods generally align, and a linear least-squares fit between them yields a slope of 1.23 $\pm$ 0.16 and an intercept of $-1.03$. 
We utilize this fitting result to convert SFR(free-free) to SFR(L$_{\mbox{\scriptsize bol}}$).
%and then to SFR(YSO). 

Here, we caution that there are potential caveats associated with the sample of massive star-forming regions used for comparison, as highlighted in previous studies \citep[e.g.,][]{Feldmann2011ApJ,Kruijssen2014MNRAS,Krumholz2019ARA&A}.
These regions are large, evolved H\textsc{ii} complexes that have likely dispersed much of their natal molecular material, meaning the current gas mass may not reliably trace the original reservoir involved in star formation.
Conversely, the stellar population may not have reached equilibrium or fully sampled the IMF in some younger regions, making SFRs derived from free-free emissions complex.
%In addition, some of these regions may still be young enough that the stellar population has not reached equilibrium or fully sampled the initial mass function
%star formation rates estimated from free-free emission may be inaccurate in these young H \textsc{ii} regions, where assumptions such as a fully sampled IMF and equilibrium conditions may not apply.

We used the N-PDF method to calculate the star-forming gas mass of these three CMZ regions, and plot them on the $M_{\rm sgb}$-SFR plot. 
They move from far below to roughly on the new \Msgb-SFR correlations. 
Thus, the new correlation may explain the origin of the very low SFR of the CMZ regions: Although a large fraction of gas is of high volume density, only a small fraction of the gas is bound, which contributes to star formation in these regions.

Our findings are consistent with both observational and theoretical predictions that the threshold for star formation is environment-dependent \citep[e.g.,][]{Kruijssen2014MNRAS,Rathborne2014ApJ,Krumholz2015MNRAS,Henshaw2023ASPC}.

\subsection{The contribution of dense, turbulent gas to star formation}

There is gravitationally dominated, dense, yet turbulent
gas in these clouds, characterized by $N >$ $N_{{\mbox{\scriptsize threshold}}}$ falling under the lognormal distribution in the N-PDF plot. 
The overall influence of these bound yet turbulent gases on star formation remains uncertain, particularly in extreme environments.  
For example, it makes up about 50\% of the total dense gas mass in the $-$1$^{\circ}$ $< \it{l} < $ 1$^{\circ}$, $\vert b \vert$ $<$ 0.5$^{\circ}$ area in the CMZ; if we remove it, this most extreme area will move even closer to the new \Msgb-SFR correlation. 
How critical is this dense and turbulent gas to star formation? 

\cite{burkhart2019} proposed that the gas components with $N >$ $N_{{\mbox{\scriptsize threshold}}}$ should be capable of collapsing to form stars, therefore categorized them as star-forming gas. 
However, there are exceptions.  
For example, some dense and turbulent gas in the Ophiuchus molecular cloud with $N >$ $N_{{\mbox{\scriptsize threshold}}}$ in N-PDF are not located within the main cloud but are isolated \citep{Jiao2022}. 
These components are not massive enough to form stars in the near future. 
Furthermore, such gas components can also reduce their density because of turbulence dissipation, moving back and forth around $N_{{\mbox{\scriptsize threshold}}}$ in the N-PDF. 
Consequently, this type of gas might only partially contribute to star formation.

We refer to the gas with $N >$ $N_{{\mbox{\scriptsize threshold}}}$ in N-PDF as a `threshold cut' gas, i.e., 
\begin{equation}\label{eq:threshcut}
    M_{\mbox{\scriptsize sgb}}^{\mbox{\scriptsize threshold cut}} = \int^{+\infty}_{N_{{\mbox{\scriptsize threshold}}}}p(N)N\mu m_{H}dN 
\end{equation}
and the bound gas without high-turbulent component that only focuses gas above the lognormal distribution in N-PDF as 'lognormal cut' gas, i.e., 
\begin{equation}\label{eq:lognormalcut}
    M_{\mbox{\scriptsize sgb}}^{\mbox{\scriptsize lognormal cut}} = \int^{+\infty}_{N_{{\mbox{\scriptsize threshold}}}}[p(N) - p_{\mbox{\scriptsize lognormal}}(N)]N\mu m_{H}dN,
\end{equation}
where $p_{\mbox{\scriptsize lognormal}}$ is the extrapolated lognormal component of the N-PDF fit.

\begin{figure}
\includegraphics[width=0.98\linewidth]{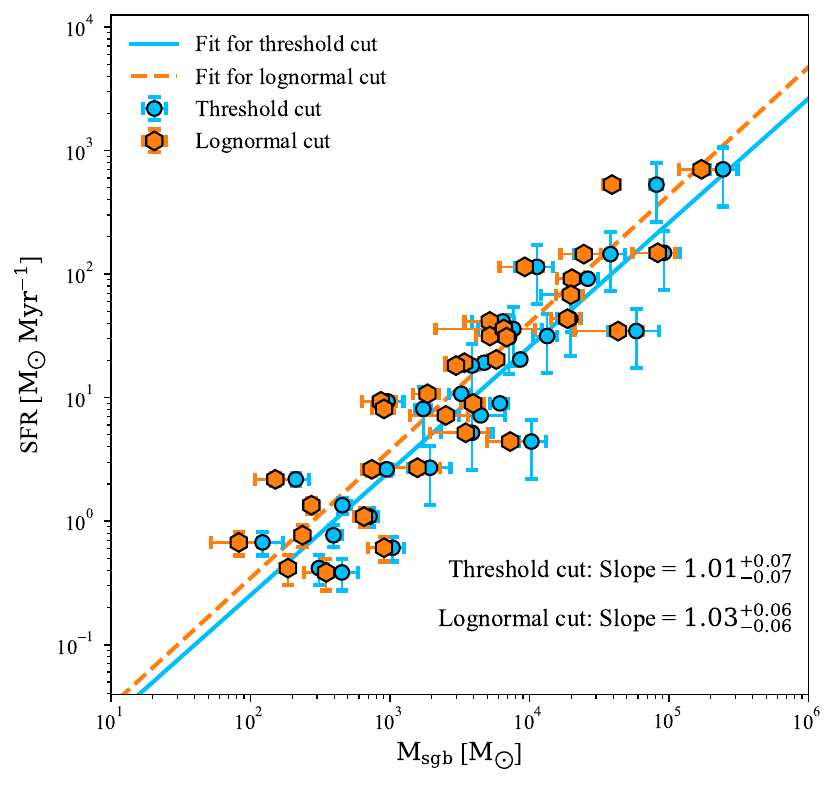}
\caption{Comparing the self-gravitating gas mass vs. SFR correlations between the N-PDF method with an unfixed column density threshold ($M_{\mbox{\scriptsize sgb}}^{\mbox{\scriptsize threshold cut}}$ Equation \ref{eq:threshcut}, the blue circles) and the N-PDF method above the lognormal distribution ($M_{\mbox{\scriptsize sgb}}^{\mbox{\scriptsize lognormal cut}}$ Equation \ref{eq:lognormalcut}, orange hexagons). }
The fitting slopes are 1.01$\pm$0.07 and 1.03$\pm$0.06, respectively. 

\label{fig:mass_sfr_compare2}
\end{figure}

We test how these two methods influence our results.
%to the one that removes any gas under the lognormal distribution from
%the N-PDF distribution, and test how these two methods may affect our results. 
The distinction between these two methods lies in whether we consider the dense, turbulent gas as star-forming gas or not.  
We apply these two approaches to calculate the mass of star-forming gas and assess the resulting $M_{\mbox{\scriptsize sgb}}$-SFR correlations.  
As demonstrated in Fig. \ref{fig:mass_sfr_compare2}, these two approaches yield very similar results.
Both show a tight, linear fit, with a slope of $\sim$1. 
Thus, although these dense but turbulent gases may contribute only partially to star formation, their inclusion or exclusion from the $M_{\mbox{\scriptsize sgb}}$ calculation does not alter our primary conclusions.

%%%%%%%%%%%%%%%%% figure 9 %%%%%%%%%%%%%%%%%%%

\section{Conclusions}\label{sec:con}

%\st{We utilize the column density probability distribution function (N-PDF) to separate gravitationally bound gas and turbulence dominated gas, and define the star forming gas based on gravitationally bound gas. We have made major conclusions below:}
%\brevise{

We analyzed the  column density probability distribution function (N-PDF) for a sample of star-forming molecular clouds that cover wide ranges of gas masses and turbulence.
Our sample comprises solar neighborhood clouds ($d<$500 pc) in which the star-formation rates (SFRs) can be inferred by YSO counts, and the distant, massive star-froming molecular clouds in which the SFRs can be inferred by infrared luminosities. 
We decomposed the N-PDF into a lognormal component in the low column density end and a power-law component in the high column density end. 
We compared the gas masses in these components with the SFRs.
Our major findings are:

%}

\begin{enumerate}
\item A surprisingly good linear correlation is found between the star formation rate and the mass of the gas that belongs to the power-law component in the N-PDF.   
\item Gas belongs to the power-law component of the N-PDF may have an universal star-forming efficiency, which is about 0.4\% per million years up to the uncertainty of converting infrared luminosity to SFR. 
\item The transitional column density between the lognormal and power-law components, $N_{\rm threshold}$, varies from cloud to cloud. In our samples, it is distributed in the range of 1--17$\times$10$^{21}$ cm$^{-2}$. Intriguingly, the $N_{\rm threshold}$ values we measured from the solar neighbor clouds approximately correspond to the extinction $A_{\rm V} = 8$. This may explain why $A_{\rm V} = 8$ has been regarded as a good empirical criterion to separate star-forming gas from the rest of the gas in a molecular cloud.
\item We applied the same analyses to three regions in the central molecular zone (CMZ). We found that in these regions, gas belongs to the power-law component of the N-PDF may have similar SFE as what were measured from outside of the CMZ.
\end{enumerate}

With these, we suggest that $N_{\rm threshold}$ is a better criterion than $A_{\rm V} = 8$ for identifying star-formation gas.
%The $A_{\rm V} = 8$ criterion has been an empirical one that has no clear physical meaning.
%Instead, theoretically, 
The $N>N_{\rm threshold}$ regions may be produced due to self-similar self-gravitational collapse (thus gravitaionally bound gas); regions at lower column densities may be supported (e.g., by gravity, magnetic field, or other feedback mechanisms) against the gravitational collapse. 
$N_{\rm threshold}$ varies from cloud to cloud may be due to an interplay between self-gravity and the support mechanisms. 
Likely, molecular clouds in the CMZ have high $N_{\rm threshold}$ values due to strong turbulence in those regions, thus using $A_{\rm V} = 8$ instead of $N_{\rm threshold}$ to identify star-forming gas for clouds in the CMZ yielded low estimates of SFE. 
Gravitationally bound gas identified by $N_{\rm threshold}$ therefore define the star forming gas in molecular clouds, with a consistent efficiency to convert gas into stars.

While the correlations identified in the present work will be empirically applicable,  our interpretation for them may be tested by the follow-up studies of gas kinematics and the energetics in the star-forming molecular clouds.

\begin{acknowledgements}
S.J. and J.W.W are supported by NSFC grant nos. 11988101 and 12041302, by the National Key R\&D Program of China No. 2023YFA1608004. J.W.W. thanks the support from the Tianchi Talent Program of Xinjiang Uygur Autonomous Region. Z-Y.Z. is supported by NSFC grant nos. 12041305, 12173016, and the science research grants from the China Manned Space Project with NOs.CMS-CSST-2021-A08 and CMS-CSST-2021-A07, and the Program for Innovative Talents, Entrepreneur in Jiangsu. NJE thanks the Astronomy Department of the University of Texas for research support. D.L.is a New Cornerstone investigator. H.B.L. is supported by the National Science and Technology Council (NSTC) of Taiwan (Grant Nos. 111-2112-M-110-022-MY3, 113-2112-M-110-022-MY3).
This research has made use of data from the {\it Herschel} Gould Belt survey (HGBS) project and the {\it Herschel} Infrared Galactic plane Survey (Hi-GAL). {\it Herschel} is an ESA space observatory with science instruments provided by European-led Principal Investigator consortia and with important participation from NASA. This research made use of the data from the Milky Way Imaging Scroll Painting (MWISP) project, which is a multi-line survey in $^{12}$CO/$^{13}$CO/C$^{18}$O along the northern galactic plane with PMO 13.7m telescope. 
\end{acknowledgements}

\bibliographystyle{aa}
\bibliography{reference}

%\clearpage
\begin{appendix}  %First appendix
\section{Contamination}

The majority of massive star-forming regions are located in the inner Galaxy, where foreground and background dust emission may contaminate {\it Herschel} images and introduce bias to the shape of the N-PDF. 
Such contamination can be categorized into two types:
a) diffuse dust emission originating from the Galactic plane; and b) additional sources with different distances along the line of sight, which cannot be distinguished by dust emission.

\subsection{Contamination from diffuse fore-/background}

The diffuse gas primarily influences the lower-density part of the N-PDF, which corresponds to the lognormal distribution. 
This could introduce bias by shifting the column density threshold to lower values and needs to be subtracted. 
Following the approach outlined in \cite{Schneider2013,chen2018}, we selected nearby regions without obvious emission as the background and subtracted it from the {\it Herschel} data.

\subsection{Contamination from additional targets along the line of sight}

To assess potential contamination from additional sources along the same line of sight, we use observational data of molecular lines that contain velocity information. 
For most of the massive star-forming regions, we accessed archival data of $^{12}$CO, $^{13}$CO, and C$^{18}$O J=1-0 from the publicly available FOREST Unbiased Galactic Plane Imaging Survey constructed with the Nobeyama 45-m telescope \citep[FUGIN,][]{Umemoto2017PASJ...69...78U}.
The data have an angular resolution of $\sim$20$''$ and a velocity resolution of 1.3 km s$^{-1}$, with a sensitivity of 0.24 K for $^{12}$CO and 0.12 K for $^{13}$CO and C$^{18}$O.
There are five massive star-forming regions (W3(OH), G9.62+0.10, G59.78+0.06, S106, and S158) that fall outside the mapping area of FUGIN.
For these clouds, we retrieved archival data of $^{12}$CO, $^{13}$CO, and C$^{18}$O J=1-0 from the Milky Way Imaging Scroll Painting (MWISP) project, observed by the Purple Mountain Observatory 14-m Delingha telescope \citep{Su2019}.
For MWISP data, the sensitivity is $\sim$0.45 K for $^{12}$CO with 0.16 km s$^{-1}$ velocity resolution and $\sim$0.25 K for $^{13}$CO and C$^{18}$O with 0.17 km s$^{-1}$ velocity resolution.

%%%%%%%%%%%%%%%%% figure A1 %%%%%%%%%%%%%%%%%%%
\begin{figure*}[!ht]
\centering
\includegraphics[width=0.95\linewidth]
{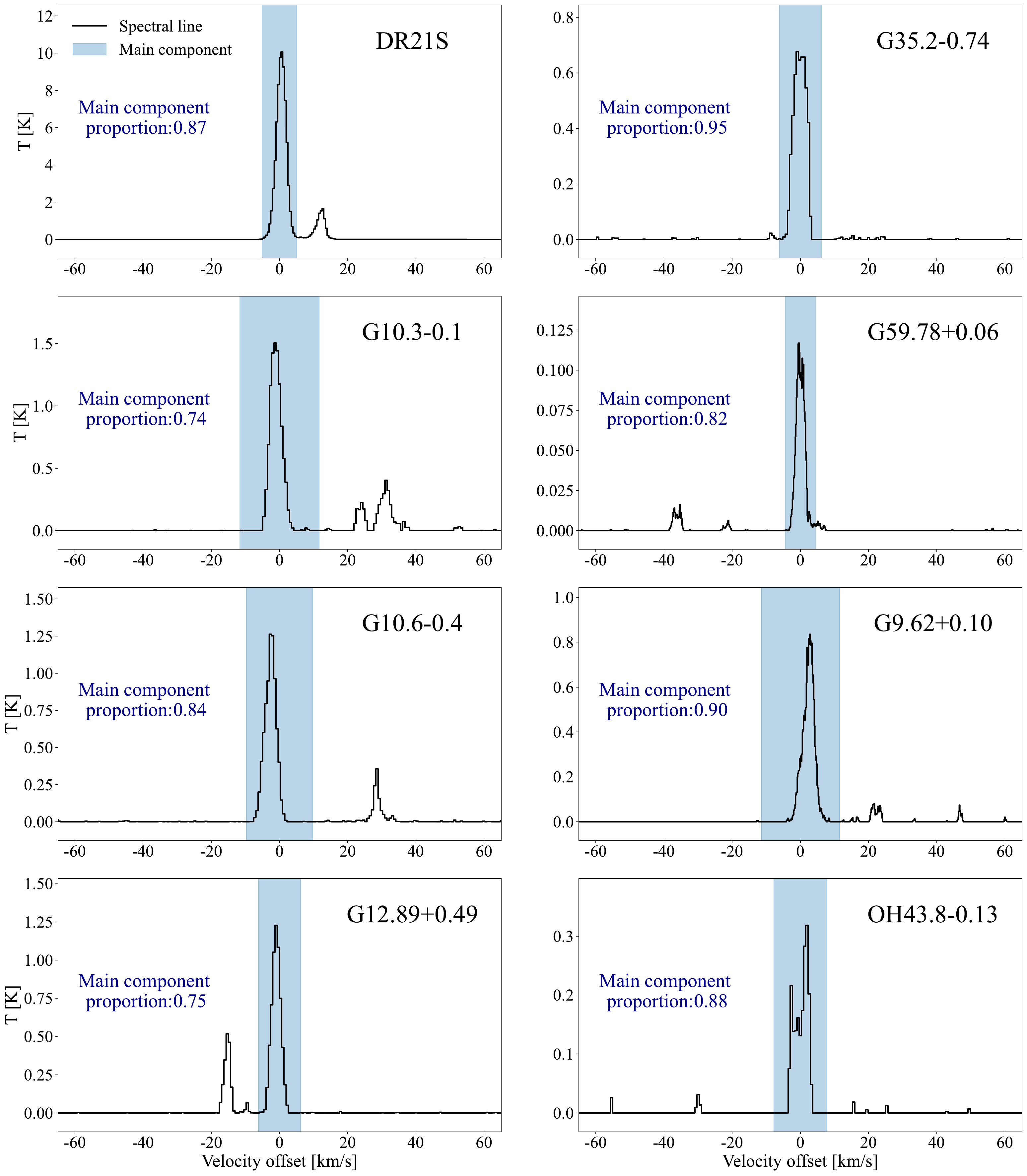}
\caption{
The C$^{18}$O spectra for DR21S, G10.6-0.4, G12.89+0.49, G35.2-0.74, G59.78+0.06, G9.62+0.10, OH43.8-0.13, S106.
The blue shadows show the velocity range of the main component.
}    
\label{fig:18co_1}
\end{figure*}
%%%%%%%%%%%%%%%%% figure A1 %%%%%%%%%%%%%%%%%%%

%%%%%%%%%%%%%%%%% figure A2 %%%%%%%%%%%%%%%%%%%
\begin{figure*}[!ht]
\centering
\includegraphics[width=0.95\linewidth]{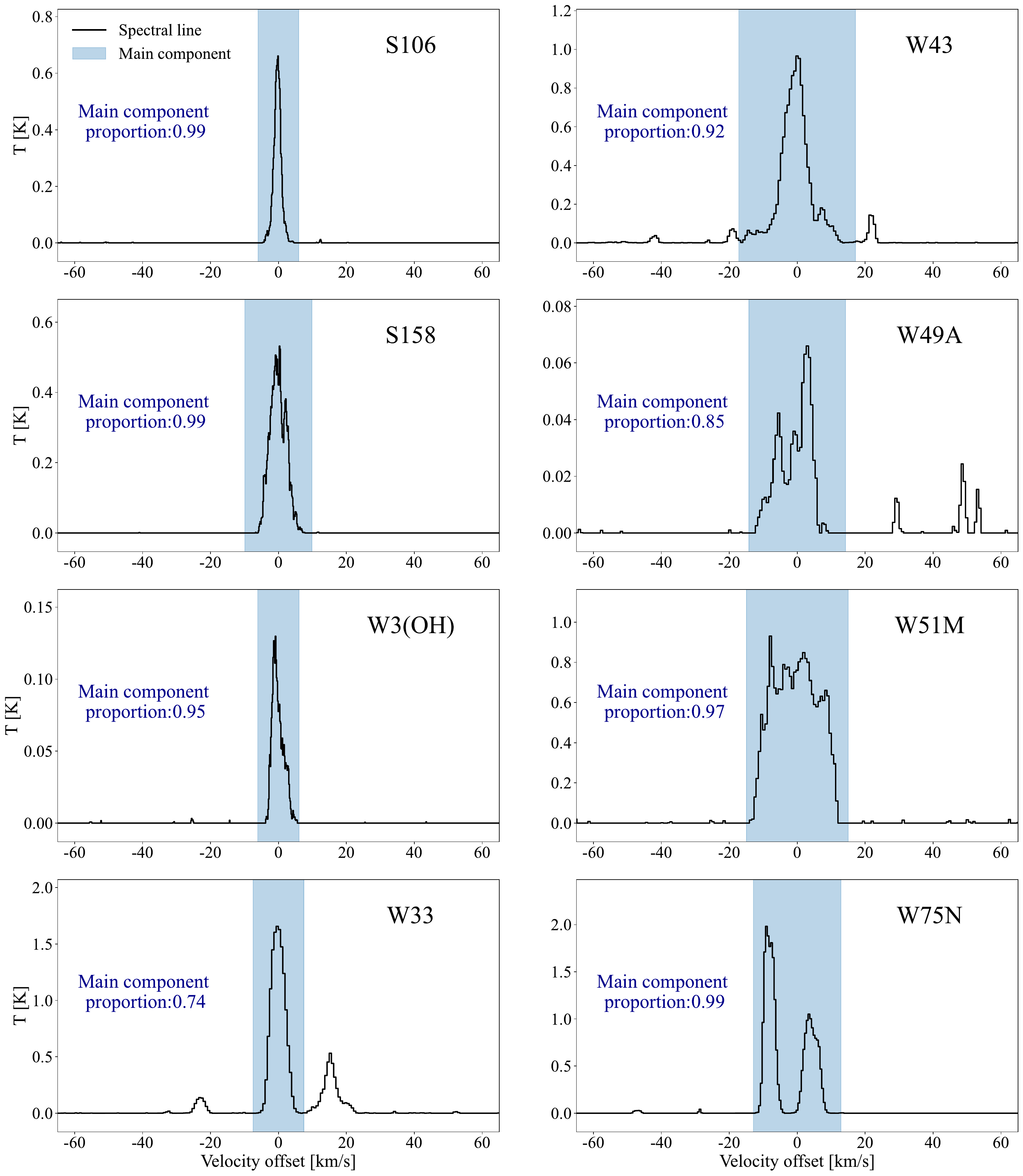}
\caption{
Similar to Fig. \ref{fig:18co_1}, but for S158, W3(OH), W31(1), W33, W43, W49A, W51M, W75N.
}    
\label{fig:18co_2}
\end{figure*}
%%%%%%%%%%%%%%%%% figure A2 %%%%%%%%%%%%%%%%%%%

Given that $^{12}$CO J=1-0 is mostly optically thick, we check the spectra of $^{13}$CO and C$^{18}$O J = 1-0 that were spatially integrated over the same region of the dust data. 
For most of the target clouds, the contribution from other velocity components is less than 20\%, whereas, for W31(1) and W33, it is $\sim$25\%.
The C$^{18}$O spectra for all targets within the gravitationally bound regions are displayed in Fig \ref{fig:18co_1} and \ref{fig:18co_2}.
Accordingly, we have incorporated a 20\% uncertainty in our estimation of \Msgb.

\section{The N-PDFs for all target clouds}

The N-PDF fitting process is described in Section \ref{subsec:npdf_fit}.
All the N-PDFs for the target clouds are attached here, and the fitting results of N-PDFs are listed in Table \ref{tab:3}.

%%%%%%%%%%%%%%%%% figure A3 %%%%%%%%%%%%%%%%%%%
\begin{figure*}[!ht]
\centering
\includegraphics[width=0.95\linewidth]{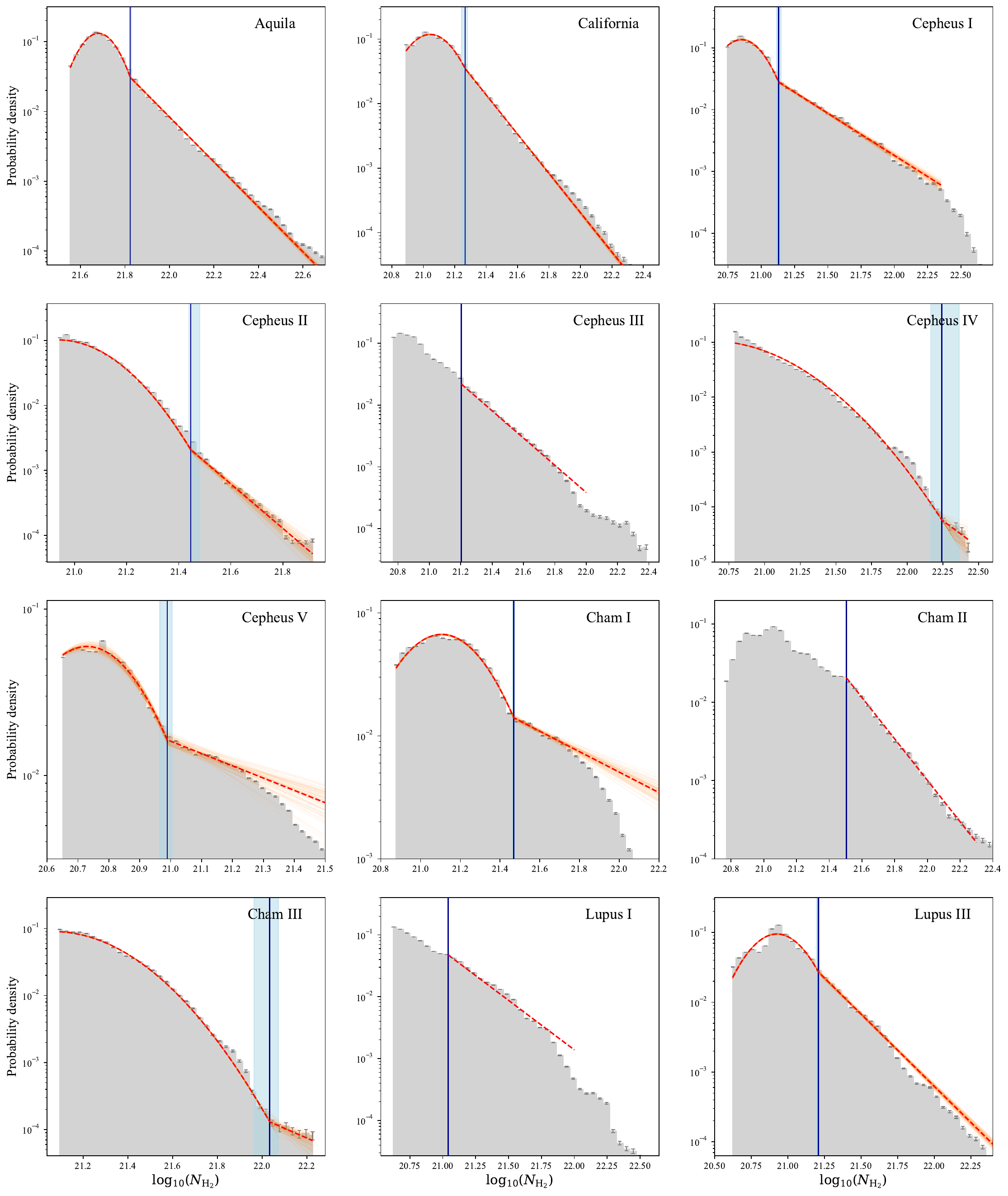}
\caption{
N-PDFs for low-mass clouds.
}    
\label{fig:lowmass_1}
\end{figure*}

%%%%%%%%%%%%%%%%% figure A4 %%%%%%%%%%%%%%%%%%%
\begin{figure*}[!ht]
\centering
\includegraphics[width=0.95\linewidth]{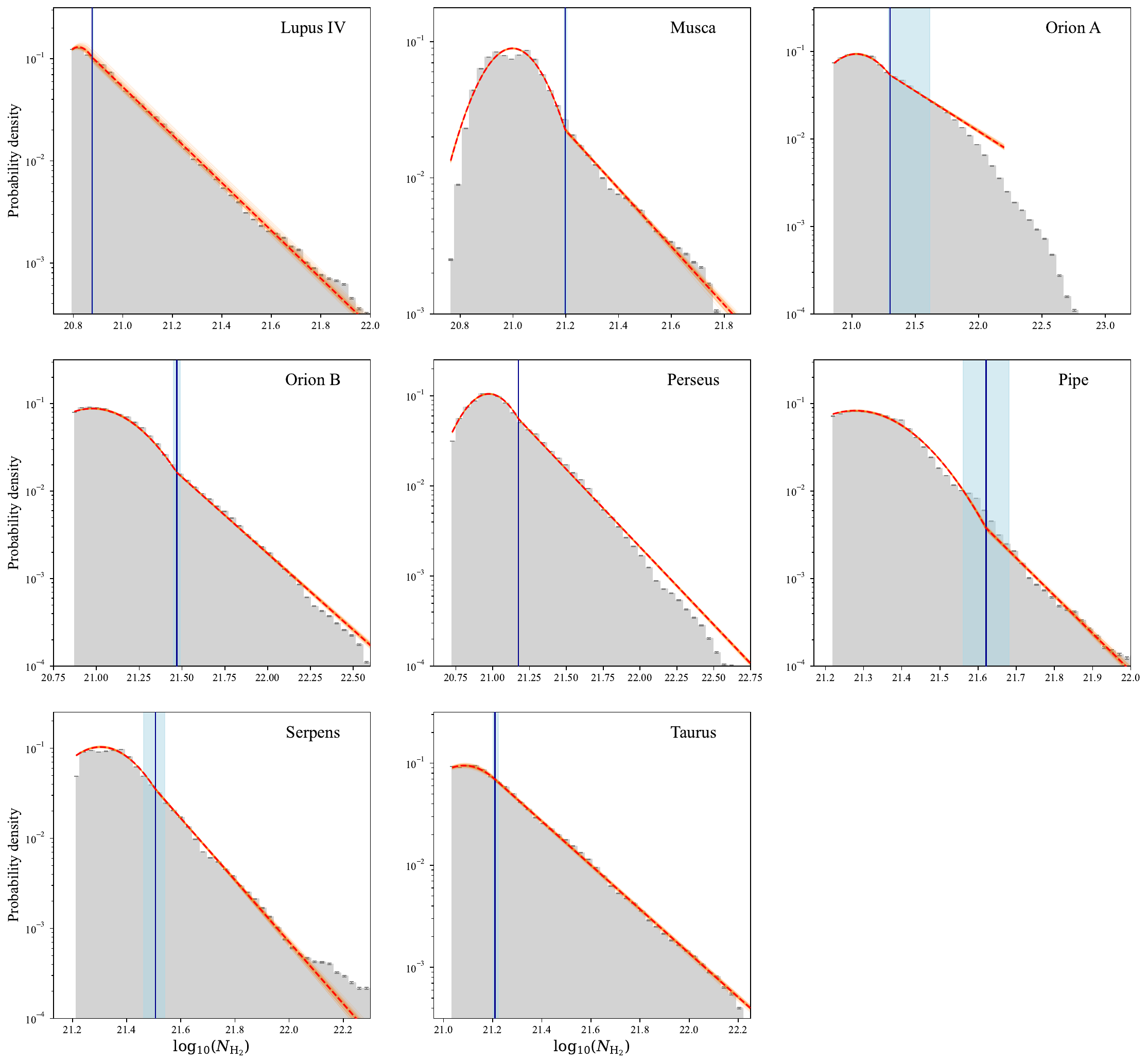}
\caption{
N-PDFs for low-mass clouds.
}    
\label{fig:lowmass_2}
\end{figure*}

%%%%%%%%%%%%%%%%% figure A5 %%%%%%%%%%%%%%%%%%%
\begin{figure*}[!ht]
\centering
\includegraphics[width=0.95\linewidth]{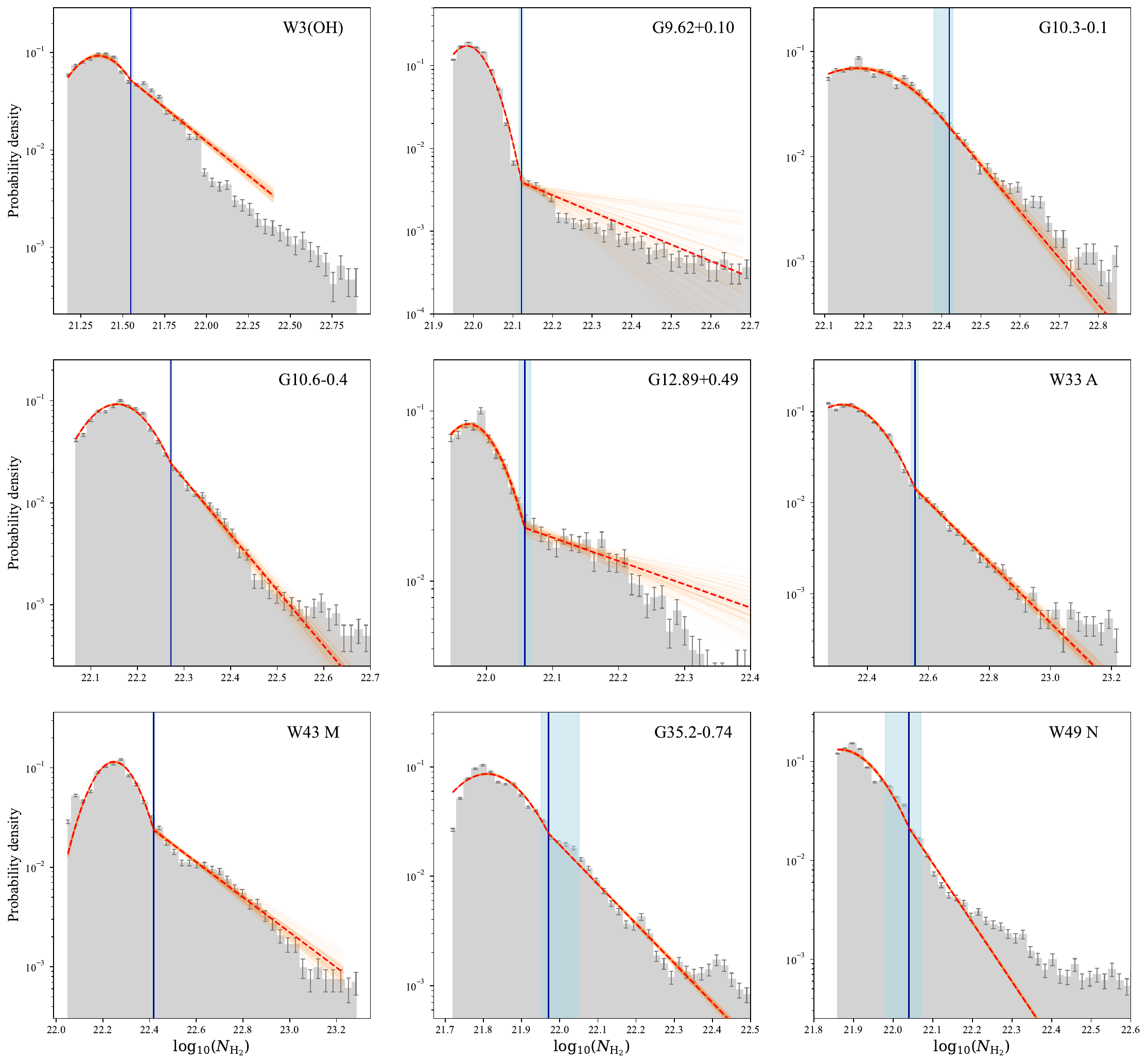}
\caption{
N-PDFs for high-mass clouds.
}    
\label{fig:highmass_1}
\end{figure*}

%%%%%%%%%%%%%%%%% figure A6 %%%%%%%%%%%%%%%%%%%
\begin{figure*}[!ht]
\centering
\includegraphics[width=0.95\linewidth]{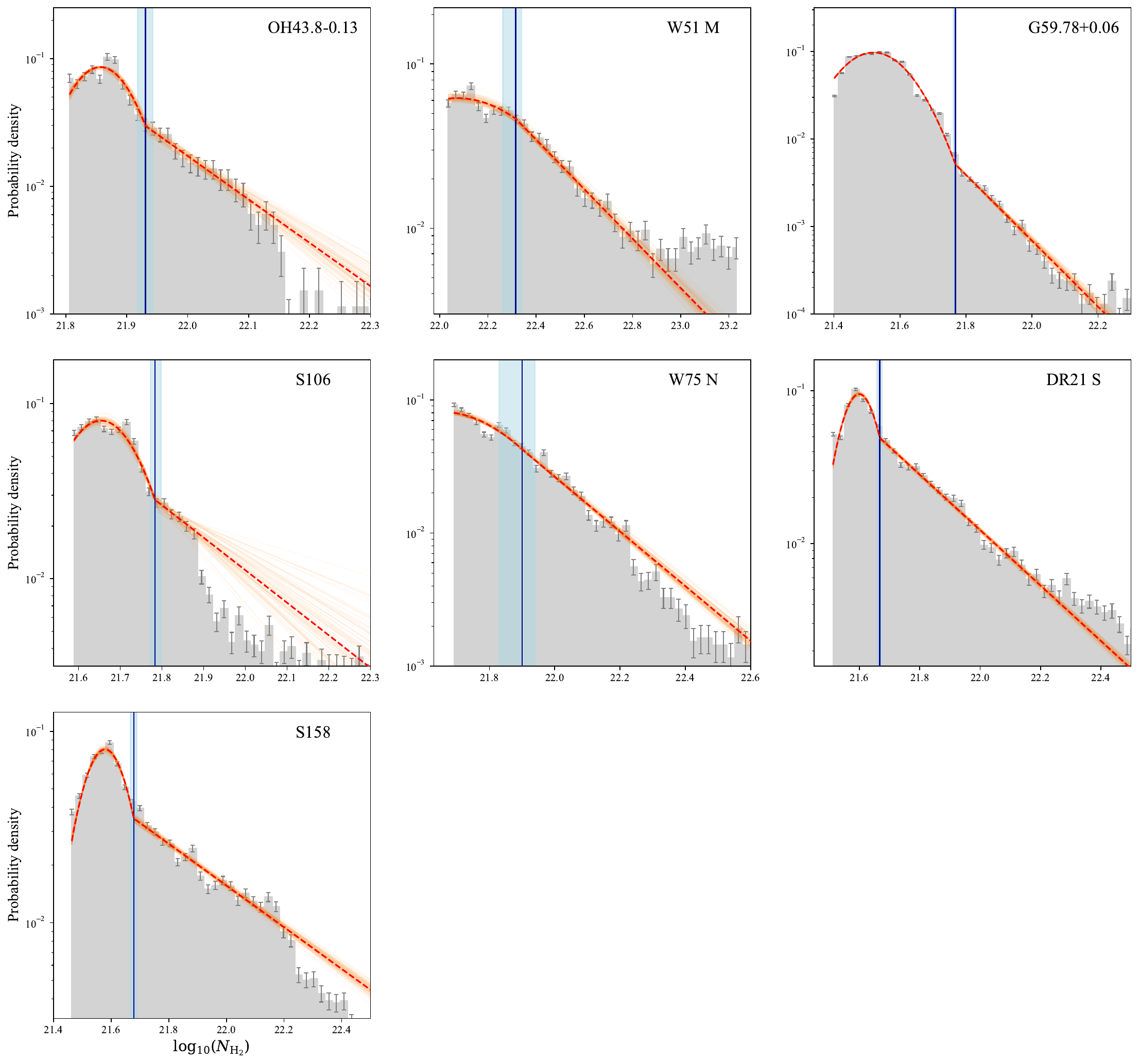}
\caption{
N-PDFs for high-mass clouds.
}    
\label{fig:highmass_2}
\end{figure*}

%%%%%%%%%%%%%%%%% table 3 %%%%%%%%%%%%%%%%%%%
\begin{table*}[!th]
\centering
\caption{Fitting results of column density probability distribution functions (N-PDFs)} \label{tab:3}
\begin{tabular}{ccccccc}
\hline
\hline \noalign{\smallskip}
Clouds & $N_{\rm cut}$ & $\langle N_{H_{2}} \rangle$ & $\eta_{t}$ & $\sigma_{\eta}$ & $\mu$ & $\alpha$ \\
  & (cm$^{-2}$) & (cm$^{-2}$) &  &  &  &  \\
\hline \noalign{\smallskip}
%------------------------------------------------------------------------
\multicolumn{7}{c}{Nearby star-forming clouds} \\ 
\hline \noalign{\smallskip}
Aquila & 3.55 $\times 10^{21}$ & 5.58 $\times 10^{21}$ & 0.18(0.01) & 0.193(0.002) & -0.151(0.002) & -3.22(0.05)  \\
California & 7.94 $\times 10^{20}$ & 1.41 $\times 10^{21}$ & 0.27(0.05) & 0.327(0.009) & -0.241(0.005) & -3.04(0.09) \\
Cepheus I & 5.25 $\times 10^{20}$  & 1.26 $\times 10^{21}$ & 0.07(0.04) & 0.361(0.021)  & -0.569(0.021) & -1.37(0.14) \\
Cepheus II & 9.33 $\times 10^{20}$ & 1.33 $\times 10^{21}$ & 0.74(0.08) & 0.422(0.006)  & -0.433(0.005) & -3.42(0.83) \\
Cepheus IV & 6.31 $\times 10^{20}$  & 1.34 $\times 10^{21}$ & 2.58(0.28)  & 0.967(0.001) & -1.170(0.001) & -1.92(0.84)   \\
Cepheus V & 4.79 $\times 10^{20}$ & 1.08 $\times 10^{21}$  & -0.10(0.05) & 0.372(0.036) & -0.697(0.046) & -0.74(0.39)  \\
Cham I & 6.34 $\times 10^{20}$ & 2.01 $\times 10^{21}$ & 0.38(0.01) & 0.472(0.003) & -0.451(0.001) & -0.83(0.10) \\
Cham III & 1.26 $\times 10^{21}$ & 2.16 $\times 10^{21}$ & 1.61(0.16) & 0.617(0.003) & -0.619(0.002) & -1.46(0.73) \\
Lupus III & 4.47 $\times 10^{20}$ & 1.14 $\times 10^{21}$ & 0.35(0.03) & 0.411(0.005) & -0.301(0.004) & -2.07(0.08) \\
Lupus IV & 5.03 $\times 10^{20}$ & 1.06 $\times 10^{21}$ & -0.35(0.01) & 0.186(0.004) & -0.475(0.003) & -2.34(0.02)  \\
Musca & 6.31 $\times 10^{20}$ & 1.25 $\times 10^{21}$ & 0.23(0.01) & 0.277(0.003) & -0.230(0.002) & -2.13(0.05) \\
Ophiuchus & 1.41 $\times 10^{21}$ & 2.74 $\times 10^{21}$ & 0.17(0.02) & 0.230(0.011) & -0.432(0.013) & -1.75(0.09)  \\
Orion A & 7.11 $\times 10^{20}$ & 2.57 $\times 10^{21}$ & -0.25(0.73) & 0.581(0.021) & -0.865(0.015) & -0.92(0.05) \\
Orion B & 7.08 $\times 10^{20}$ & 1.92 $\times 10^{21}$ & 0.43(0.05) & 0.618(0.019) & -0.703(0.030) & -1.75(0.07)  \\
Perseus & 5.62 $\times 10^{20}$ &  1.58 $\times 10^{21}$ & -0.05(0.01) & 0.409(0.003) & -0.517(0.002) & -1.72(0.01) \\
Pipe & 1.58 $\times 10^{21}$ &  2.40 $\times 10^{21}$ & 0.56(0.14) & 0.317(0.003) & -0.233(0.004) & -4.31(0.16)  \\
Serpens & 1.74 $\times 10^{21}$ &  2.77 $\times 10^{21}$ & 0.15(0.10) & 0.317(0.013) & -0.317(0.007) & -3.43(0.17) \\
Taurus & 1.05 $\times 10^{21}$ & 2.01 $\times 10^{21}$ & -0.22(0.03) & 0.367(0.009) & -0.508(0.006) & -2.16(0.01) \\
\hline \noalign{\smallskip}
\multicolumn{7}{c}{Massive star-forming regions} \\
\hline \noalign{\smallskip}
W3(OH) & 1.58 $\times 10^{21}$ & 4.09 $\times 10^{21}$ & -0.15(0.02) & 0.415(0.033) & -0.593(0.022) & -1.39(0.16)  \\
G9.62+0.10 & 8.91 $\times 10^{21}$ & 1.07 $\times 10^{22}$ & 0.21(0.01) & 0.115(0.001) & -0.109(0.002) & -1.99(0.49)  \\
G10.30-0.10 & 1.35 $\times 10^{22}$ & 2.02 $\times 10^{22}$ & 0.26(0.09) & 0.338(0.024) & -0.280(0.033) & -4.43(0.68)  \\
G10.60-0.40 & 1.20 $\times 10^{22}$ & 1.63 $\times 10^{22}$ & 0.14(0.01) & 0.163(0.002) & -0.127(0.001) & -5.44(0.12)  \\
G12.89+0.49 & 8.91 $\times 10^{21}$ & 1.27 $\times 10^{22}$ & -0.10(0.02) & 0.117(0.008) & -0.302(0.015) & -1.37(0.87)  \\
W33A & 1.86 $\times 10^{22}$ & 2.64 $\times 10^{22}$ & 0.31(0.03) & 0.267(0.011) & -0.243(0.016) & -3.32(0.32)  \\
W43S & 1.26 $\times 10^{22}$ & 2.36 $\times 10^{22}$ & 0.10(0.01) & 0.220(0.004) & -0.289(0.003) & -1.76(0.05)  \\
G35.20-0.74 & 5.37 $\times 10^{21}$ & 8.16 $\times 10^{21}$ & 0.14(0.18) & 0.235(0.008) & -0.235(0.008) & -3.60(0.17)  \\
W49N & 7.08 $\times 10^{21}$ & 9.45 $\times 10^{21}$ & 0.15(0.14) & 0.212(0.008) & -0.256(0.015) & -5.98(0.06)  \\
OH43.80-0.13 & 6.61 $\times 10^{21}$ & 8.72 $\times 10^{21}$ & -0.02(0.03) & 0.117(0.008) & -0.193(0.016) & -3.40(1.09)  \\
W51M & 1.12 $\times 10^{22}$ & 3.30 $\times 10^{22}$ & -0.47(0.12) & 0.739(0.243) & -1.032(0.082) & -1.51(0.19)  \\
G59.78+0.06 & 2.57 $\times 10^{21}$ & 3.70 $\times 10^{21}$ & 0.46(0.02) & 0.235(0.003) & -0.111(0.004) & -3.76(0.45)  \\
S106 & 3.98 $\times 10^{21}$ & 6.65 $\times 10^{21}$ & -0.09(0.04) & 0.207(0.023) & -0.389(0.021) & -1.86(1.05)  \\
W75N & 4.79 $\times 10^{21}$ & 8.99 $\times 10^{21}$ & -0.12(0.16) & 0.506(0.037) & -0.701(0.036) & -2.06(0.18)  \\
DR21 S & 3.31 $\times 10^{21}$ & 9.20 $\times 10^{21}$ & -0.68(0.02) & 0.137(0.010) & -0.840(0.007) & -1.81(0.10)  \\
S158 & 3.09 $\times 10^{21}$ & 7.72 $\times 10^{21}$ & -0.48(0.03) & 0.179(0.014) & -0.712(0.009) & -1.09(0.09)  \\
\hline \\
\end{tabular}
\end{table*}
%%%%%%%%%%%%%%%%% table 1 %%%%%%%%%%%%%%%%%%%

\section{The effects of spatial resolution on $N_{{\mbox{\scriptsize threshold}}}$ fitting}\label{Appendix:resolution}

Since N-PDFs are simplified 1D representations of complex 2D column density structures, the influence of spatial resolution on the shape of N-PDF is not straightforward.
\citet{Alves2017} examined the N-PDFs of the Ophiuchus cloud at different spatial resolutions and found that the power-law slope remained largely unchanged as resolution decreased.

In this study, the {\it Herschel} observations have different spatial resolutions for the nearby cloud sample and the more distant, massive cloud sample due to their varying distances.
To assess whether this spatial resolution discrepancy introduces systematic differences in the fitted transitional column density ($N_{\mbox{\scriptsize threshold}}$) and the resulting $M_{\rm sgb}$ measurements, we performed a test using Ophiuchus as an example.

\begin{figure*}
\begin{tabular}{p{0.3\linewidth}p{0.3\linewidth}p{0.3\linewidth} }
\hspace{-0.3cm}\includegraphics[scale=0.33]{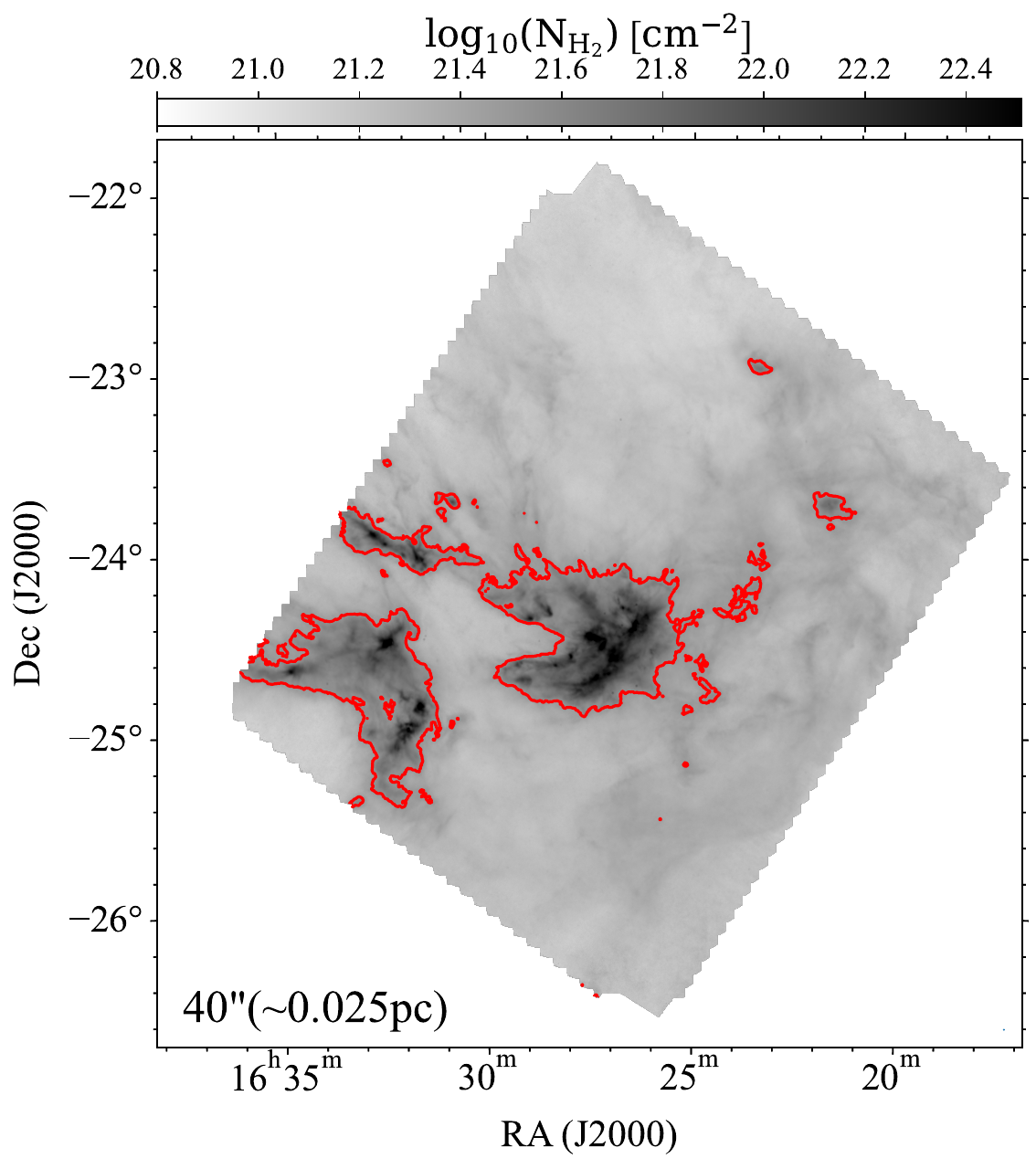} &
\hspace{0.1cm}\includegraphics[scale=0.33]{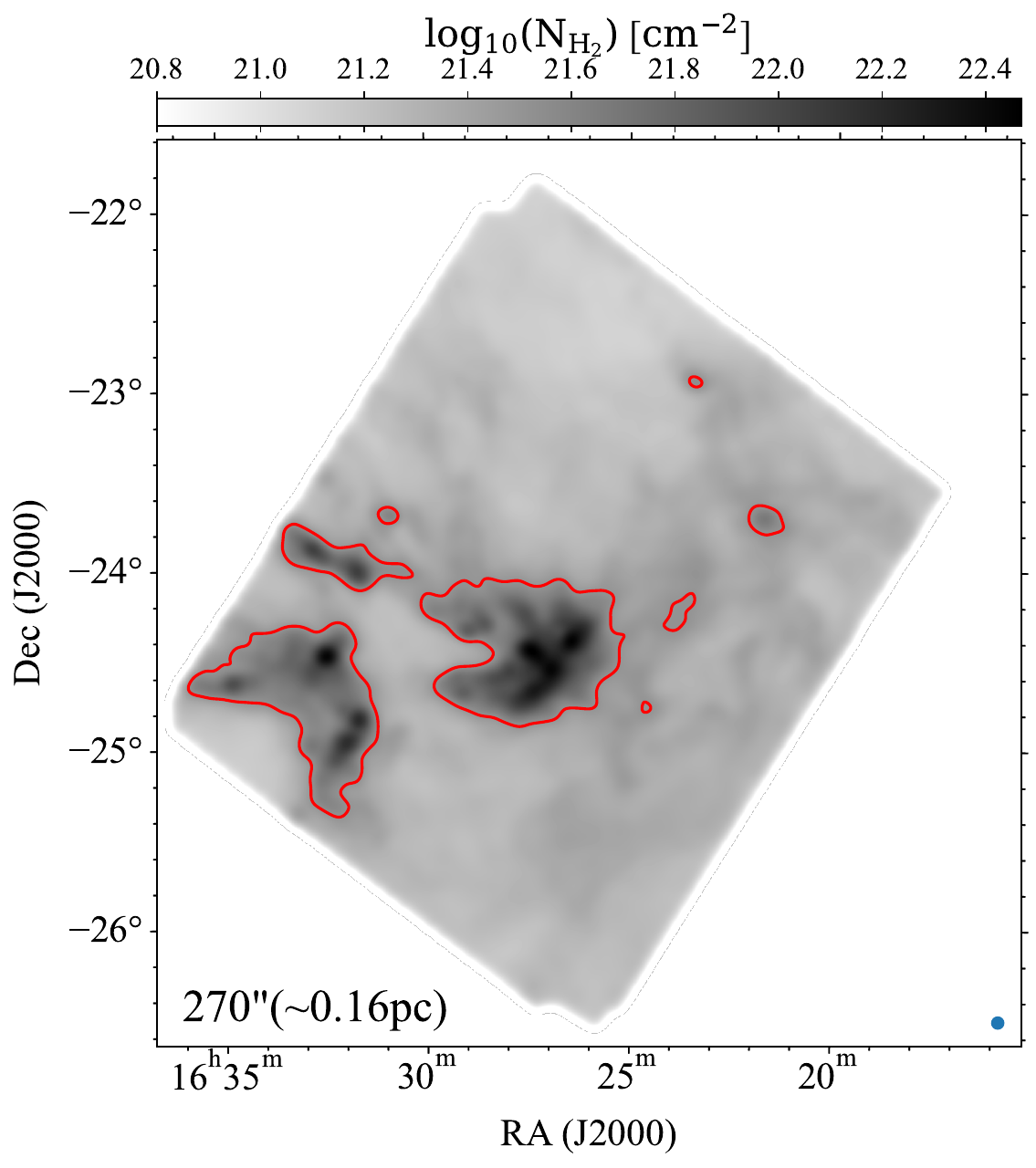}&
\hspace{0.5cm}\includegraphics[scale=0.33]{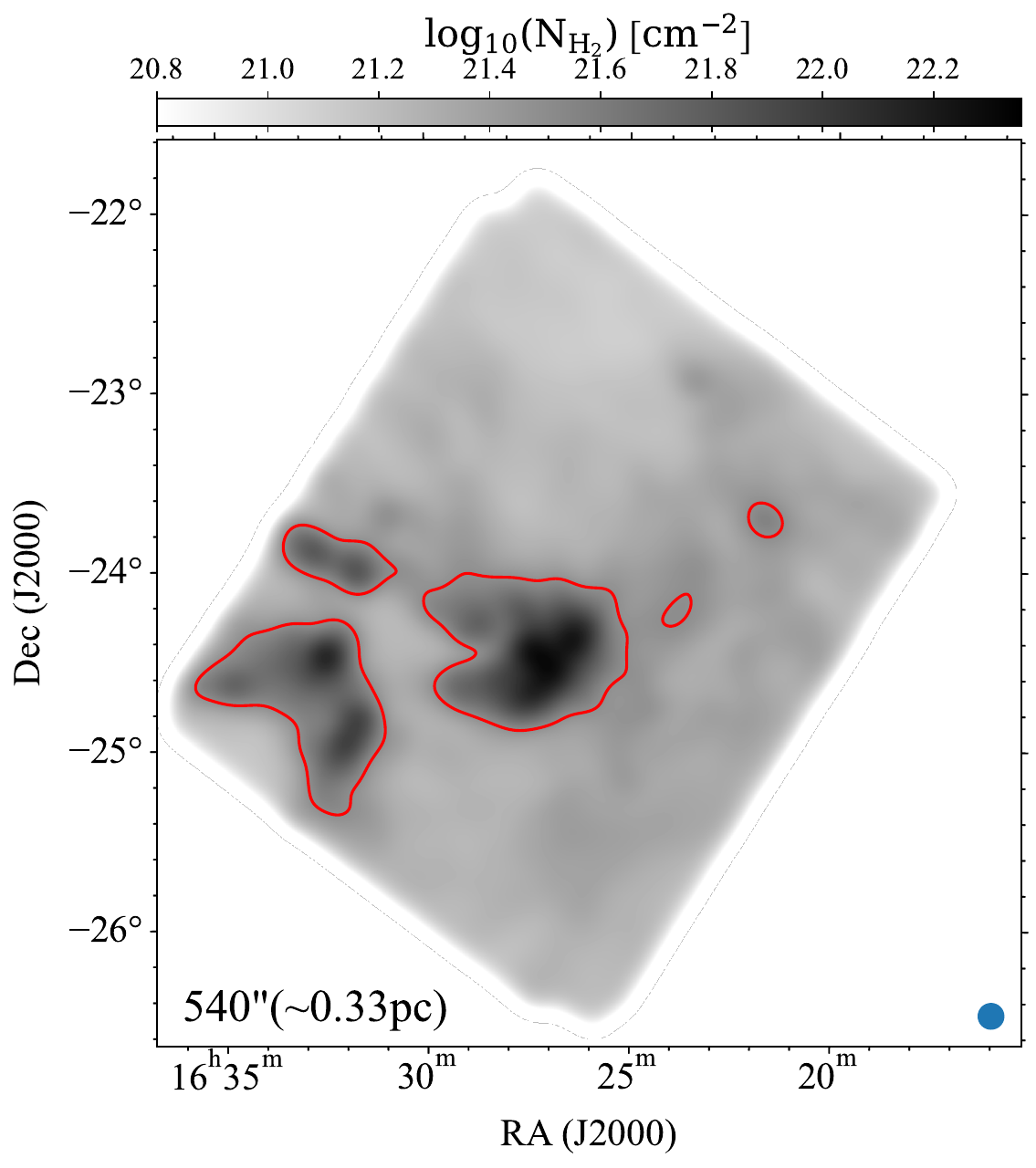}  \\
\end{tabular}
\begin{tabular}{p{0.3\linewidth}p{0.3\linewidth}p{0.3\linewidth} }
\hspace{-0.3cm}\includegraphics[scale=0.33]{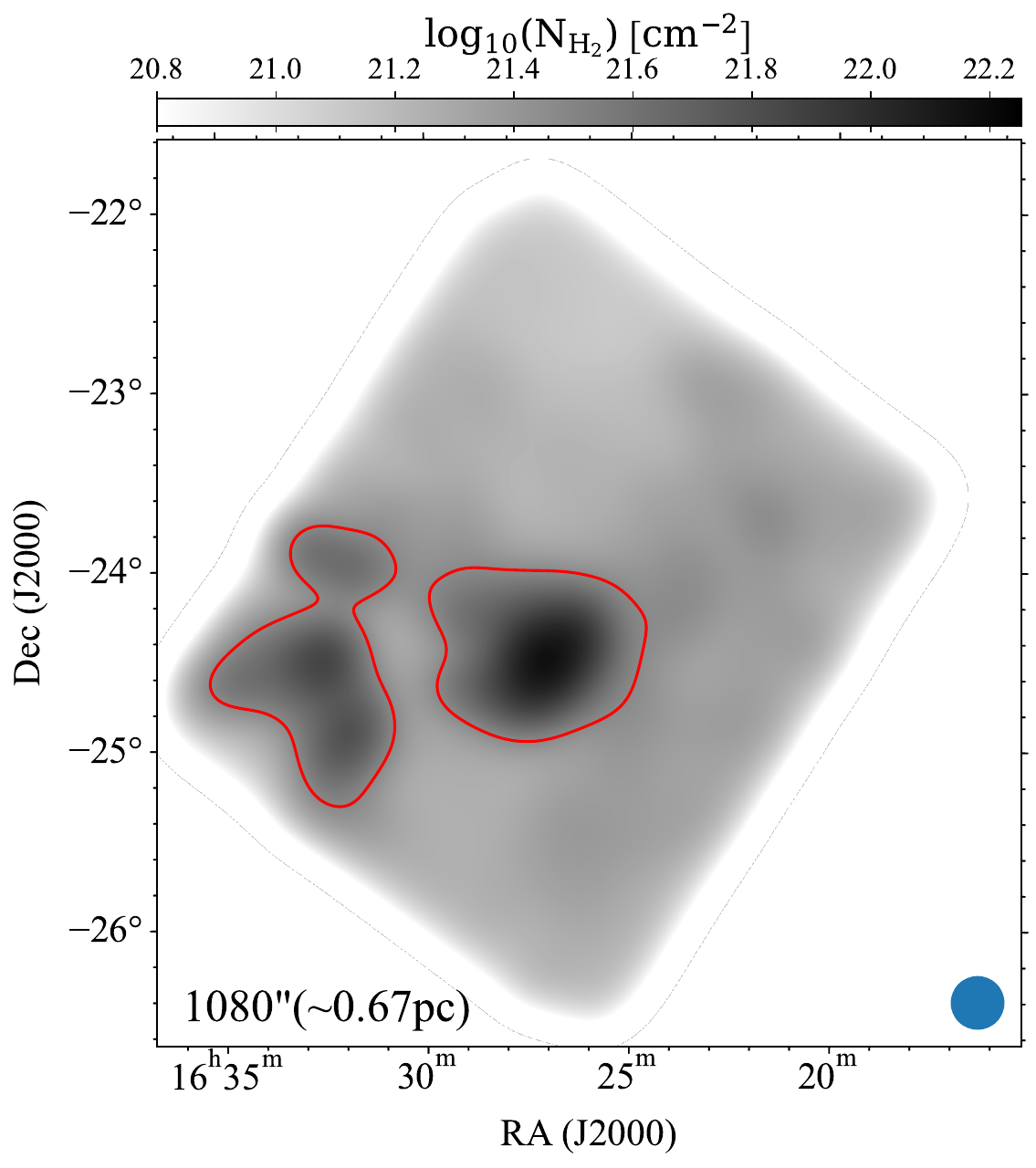} &
\hspace{0.1cm}\includegraphics[scale=0.33]{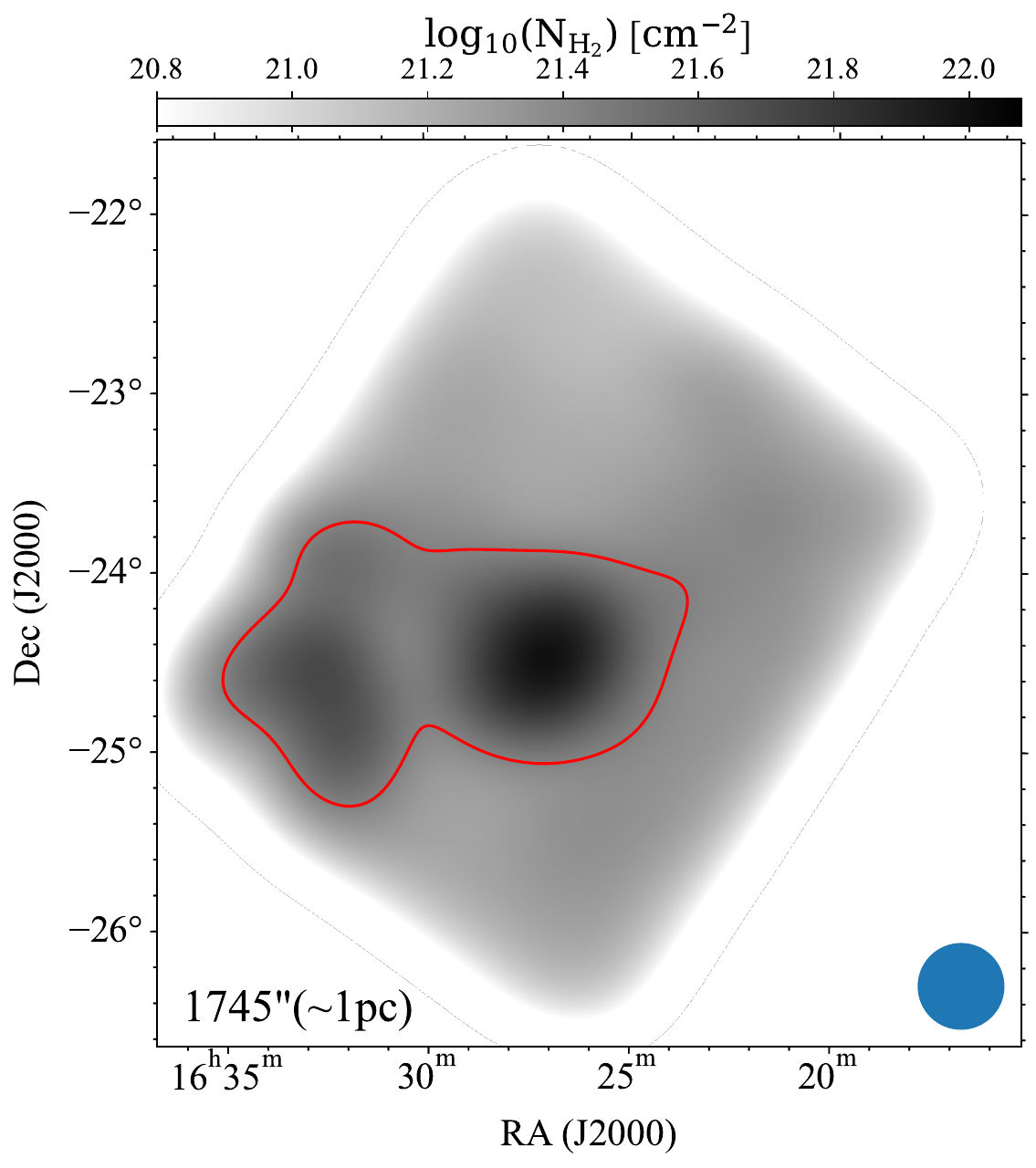}&
\\
\end{tabular}
\caption{
Column density maps of the Ophiuchus molecular cloud at different spatial resolutions.
The red contours indicate the absolute transitional column density, $N_{{\mbox{\scriptsize threshold}}}$.
}
\label{fig:resolution_N}
\end{figure*}

\begin{figure*}
\begin{tabular}{p{0.3\linewidth}p{0.3\linewidth}p{0.3\linewidth} }
\hspace{-0.4cm}\includegraphics[scale=0.42]{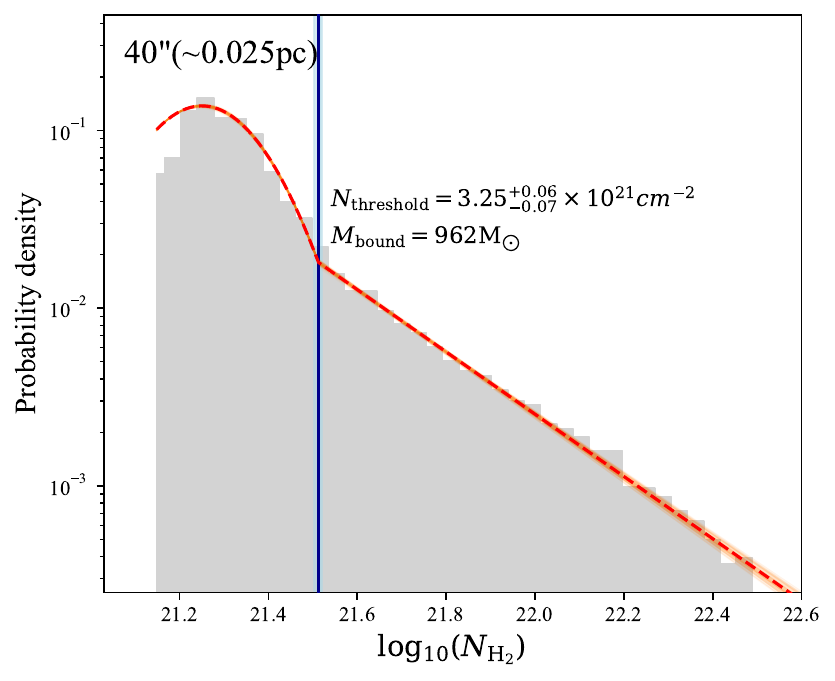} &
\hspace{0.cm}\includegraphics[scale=0.42]{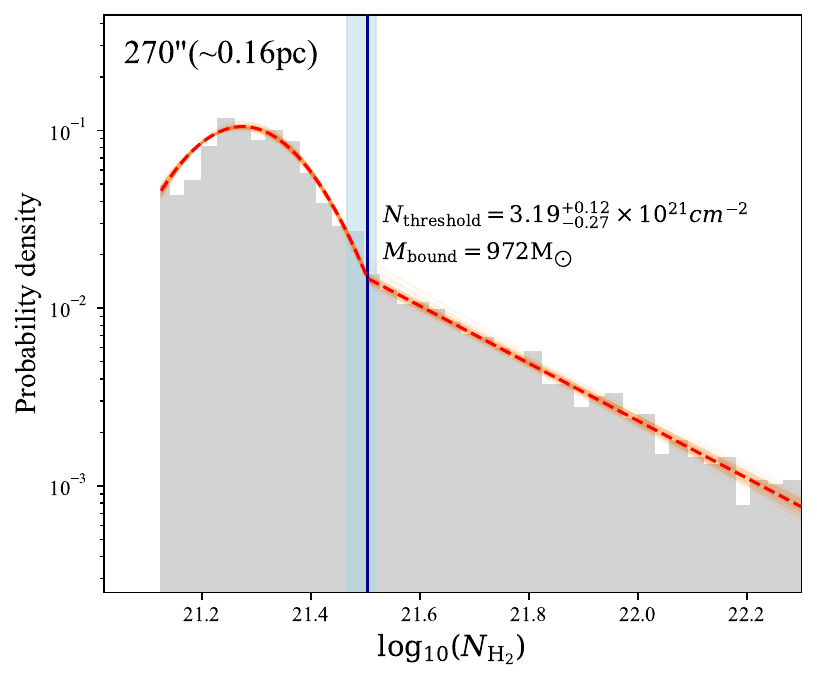}&
\hspace{0.3cm}\includegraphics[scale=0.42]{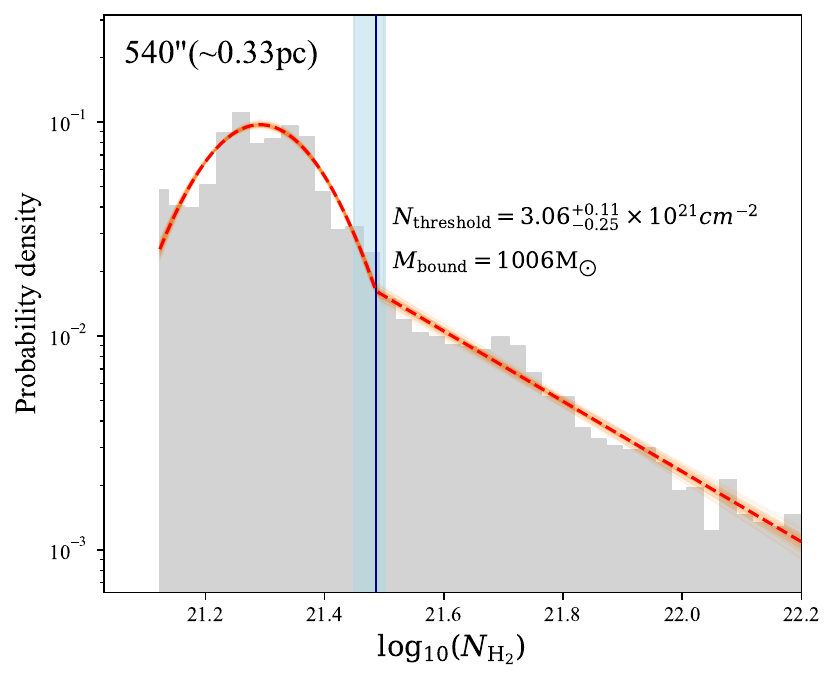}  \\
\end{tabular}
\begin{tabular}{p{0.3\linewidth}p{0.3\linewidth}p{0.3\linewidth} }
\hspace{-0.4cm}\includegraphics[scale=0.42]{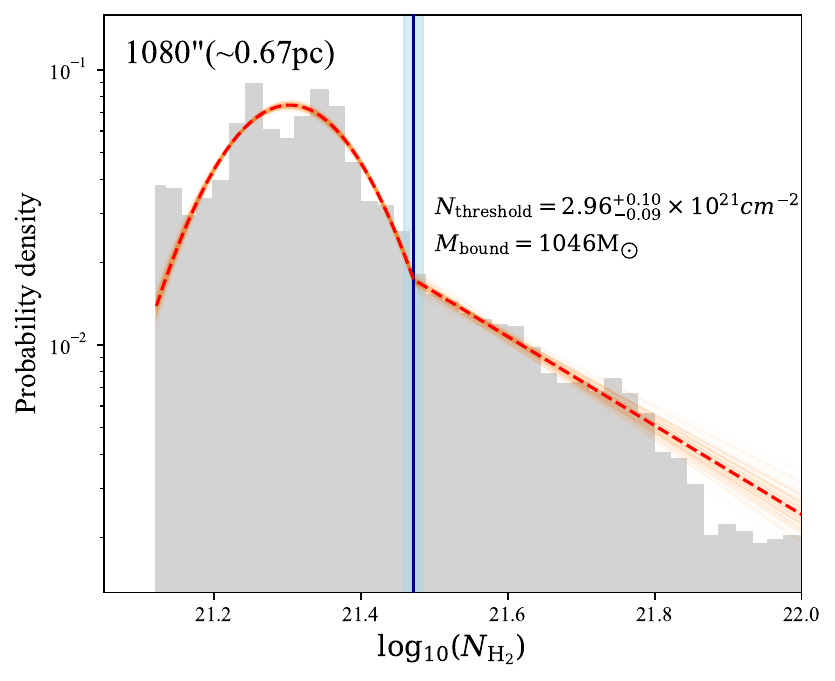} &
\hspace{0.cm}\includegraphics[scale=0.42]{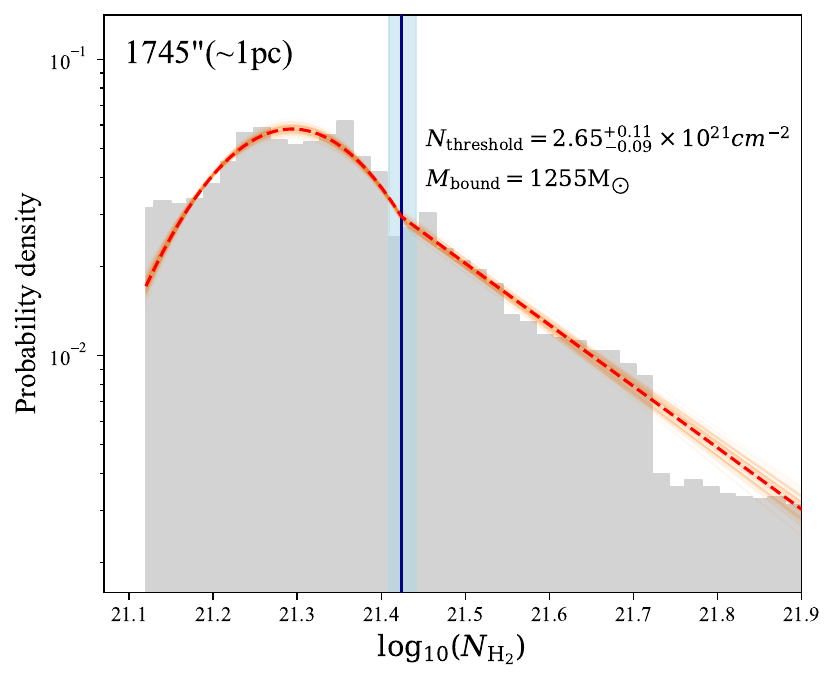}&
\\
\end{tabular}
\caption{
N-PDFs of the Ophiuchus molecular cloud at different spatial resolutions.
}
\label{fig:resolution_PDF}
\end{figure*}

The typical distance of the high-mass cloud sample is $\sim$ 4 kpc.
We therefore scaled the Ophiuchus {\it Herschel} observations to simulate distances of 1, 2, 4, and 6 kpc, yielding column density maps using SED fitting with spatial resolutions of $\sim$ 0.15, 0.30, 0.60, and 1.00 pc, respectively.
%By scaling the Ophiuchus {\it Herschel} observations to the distances of 1, 2, 4, and 6 kpc, we generated the column density maps of the Ophiuchus molecular cloud based on SED fitting, each with spatial resolutions of $\sim$ 0.15, 0.30, 0.60, and 1.00 pc, respectively.
Subsequently, we analyze the N-PDFs of the Ophiuchus molecular at different spatial resolutions.
Figure \ref{fig:resolution_N} shows the column density maps at different spatial resolutions, while Figure \ref{fig:resolution_PDF} presents the corresponding N-PDFs.
The red contours in Figure \ref{fig:resolution_N} mark the transitional column density from the N-PDFs.
Notably, in the N-PDFs at different spatial resolutions, the high-density end decreases as the spatial resolution decreases, while the transitional column density and the gravity-bound structure remain essentially invariant.
Moreover, the variation in the estimated bound gas mass is within 30\%.
This implies that the transitional column density of N-PDF is invariant to spatial resolution once the spatial resolution can resolve the gravity-bound structures.

%%%%%%%%%%%%%%figure 1%%%%%%%%%%%%%%%

%%%%%%%%%%%%%%%%%%%%%%%%%%%%%%%%%%%%

%%%%%%%%%%%%%%figure 2%%%%%%%%%%%%%%%

%%%%%%%%%%%%%%%%%%%%%%%%%%%%%%%%%%%%

\end{appendix}

\end{document}